\documentclass[12pt]{article}

\usepackage{graphicx}
\usepackage[letterpaper,margin=1in]{geometry}
\renewenvironment{abstract}
	{\quotation}
	{\endquotation}

\date{}

\makeatletter
\renewcommand{\fnum@figure}{\textbf{Figure \thefigure}}
\renewcommand{\fnum@table}{\textbf{Table \thetable}}
\makeatother

\usepackage{url}

\usepackage{braket}
\usepackage{nicefrac}
\usepackage{amsmath}
\usepackage[hidelinks]{hyperref}
\usepackage{rotating}
\usepackage{color}

\newcommand{\MnSn}{Mn$_{3}$Sn}
\newcommand{\MnGe}{Mn$_{3}$Ge}
\newcommand{\MnGa}{Mn$_{3}$Ga}
\newcommand{\MnX}{Mn$_{3}X$ ($X=$~Ga, Ge, Sn)}
\newcommand{\MnXGa}{Mn$_{x}$Ga}
\newcommand{\hexsg}{$P6_{3}/mmc$ (No.~194)}
\newcommand{\monosg}{$P2_{1}/m$ (No.~11)}
\newcommand{\GMT}{$\Gamma_{2}^{+}$}
\newcommand{\GMF}{$\Gamma_{5}^{+}$}
\newcommand{\MnO}{MnO}

\newcommand{\magsghrpt}{$P6_{3'}/m'mc'$ (BNS 194.269)}
\newcommand{\magsgh}{$Cm'cm'$ (BNS 63.464)}
\newcommand{\magsgl}{$P2_{1'}/m'$ (BNS 11.54)}
\newcommand{\orthosg}{$Cmcm$ (No.~63)}

\title{Intrinsic Topological Weyl Phase Transition Induced by a Magnetostructural Transformation in a Kagome Magnet}

\author{
  Tsung-Han Yang$^{1}$,
  Satoshi Okamoto$^{2,\ast}$,
  D. Alan Tennant$^{3,4,5}$,\\
  Michael A. McGuire$^{2,\ast}$,
  Qiang Zhang$^{1,\ast}$\\
  \small $^{1}$ Neutron Scattering Division, Oak Ridge National Laboratory,\\ 
  \small Oak Ridge, Tennessee 37831, USA\\
  \small $^{2}$ Materials Science \& Technology Division, Oak Ridge National Laboratory,\\ 
  \small Oak Ridge, Tennessee 37831, USA\\
  \small $^{3}$ Department of Physics \& Astronomy, University of Tennessee, Knoxville, TN, USA\\
  \small $^{4}$ Shull Wollan Center, Oak Ridge National Laboratory, Oak Ridge, TN 37831, USA\\
  \small $^{5}$ Department of Materials Science and Engineering, University of Tennessee, Knoxville, TN, USA\\[6pt]
  \small $^{\ast}$ Corresponding author: \texttt{okapon@ornl.gov}\\
  \small $^{\ast}$ Corresponding author: \texttt{mcguirema@ornl.gov}\\
  \small $^{\ast}$ Corresponding author: \texttt{zhangq6@ornl.gov}
}

\begin{document}

\maketitle

\begin{abstract} \bfseries \boldmath

Topological phase transitions provide a unique window into the interplay between structure, magnetism, and Weyl physics in magnetic Weyl semimetals. However, realizing an intrinsic Weyl phase transition between two distinct Weyl states near room temperature remains challenging. Here, we demonstrate that a magnetostructural transition effectively induces such a transition in the kagome magnet \MnGa{}. 
High-resolution neutron diffraction, magnetization characterizations and first-principles calculations reveal that \MnGa{} undergoes a chiral antiferromagnetic transition below 485~K, followed by a magnetostructural transition to a monoclinic structure with highly canted antiferromagnetic order near room temperature. 
These cooperative changes in lattice and magnetic symmetries reorganize Weyl nodes, driving a transition from a primary type-II Weyl state to a distinct Weyl state, accompanied by dramatic variations in the anomalous Hall effect and appearance of topological Hall effect. Our findings open a new pathway for discovering novel topological Weyl states and  potential spintronic applications. 

\end{abstract}

\noindent
\section{Introduction}
The discovery of topological quantum materials has led to transformative insights into condensed matter physics, with Weyl semimetals emerging as a particularly intriguing class due to their gapless excitations associated with Weyl nodes, band crossings which act as monopoles of Berry curvature in momentum space~\cite{Weng2015,Wan2011,Xu2015,Lu2015,Lv2015,Armitage2018}. 
While early realizations focused on nonmagnetic systems with broken inversion symmetry,  magnetic Weyl semimetals have recently garnered significant interest, as the time-reversal symmetry breaking by magnetic order offers additional control over the topological band structure~\cite{Nakatsuji2015,Wang2018,Liu2019,Zou2019}. Exploring topological phase transitions provides an excellent  opportunity to investigate the intricate coupling among structure, magnetism, and Weyl physics. To date, only a limited number of compounds have been identified to undergo topological phase transitions from one non-Weyl topological state to a Weyl state, primarily driven by external stimuli such as magnetic fields or pressure~\cite{Lee2021,Cheng2024,Du2022}. 
For example, an applied magnetic field induces the transformation from an antiferromagnetic (AFM) topological insulator to an ideal type-II Weyl state in Mn(Bi$_{1-x}$Sb$_x$)$_2$Te$_4$~\cite{Lee2021}. 
A pressure-induced transition from a magnetic topological insulator to a trivial insulator, and then to a Weyl semimetal was reported in EuCd$_{2}$As$_{2}$~\cite{Du2022}. 
From the perspective of both fundamental and applied research, it is particularly intriguing yet challenging to explore strategies for inducing intrinsic topological phase transitions between two distinct Weyl states without external stimuli, especially under conditions near room temperature. 

\begin{figure}
    \centering
    \includegraphics[width=0.55\linewidth]{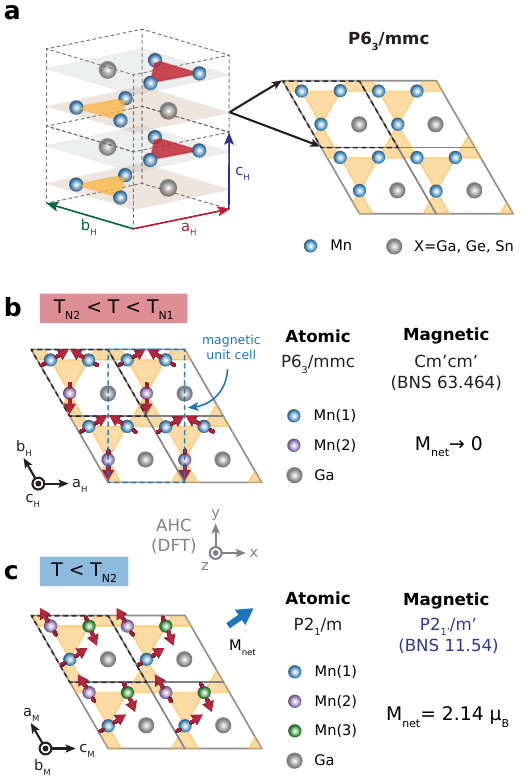}
\caption{
\textbf{Crystal and magnetic structures of \MnGa{}.
} \textbf{a} The high-temperature paramagnetic phase with hexagonal \hexsg{} symmetry, isostructural to \MnSn{} and \MnGe{}. Mn atoms form two-dimensional kagome layers stacked along the hexagonal $c_\text{H}$ axis. \textbf{b} Chiral AFM magnetic structure in the intermediate-temperature phase ($T_\text{N2}<T<T_\text{N1}$). While the underlying crystal structure remains a hexagonal structure, magnetic ordering breaks the six-fold rotational symmetry, resulting in an orthorhombic magnetic unit cell (indicated by the blue dashed line), with magnetic space group \magsgh{}. \textbf{c} Low-temperature phase ($T<T_\text{N2}$). The crystal symmetry lowers to monoclinic \monosg{}, accompanied by a reorientation of Mn magnetic moments. The resulting magnetic space group is \magsgl{}, with a large net ferromagnetic moment of $\textbf{M}_\text{net}\approx$2.14~$\mu_{B}$ oriented approximately perpendicular to the $a_\text{M}$ axis. The gray axes ($x$, $y$, $z$) define the laboratory frame used to calculate AHC in both phases.}
\label{fig:schematics}
\end{figure}

The kagome magnets \MnSn{} and \MnGe{}, which share the same hexagonal\textbf{} structure as \MnGa{} shown in Fig.~\ref{fig:schematics}a, have attracted significant attention~\cite{Nakatsuji2015,Kuebler2014,Nayak2016,Ikhlas2017,Yang2017,Kuroda2017,Kimata2019,Tsai2020} due to their recent identification as magnetic Weyl semimetals exhibiting chiral magnetic order and remarkable quantum transport properties.
In \MnSn{}, the interplay between the noncollinear antiferromagnetic order and strong spin-orbit coupling gives rise to significant anomalous and topological Hall effects (AHE and THE)~\cite{Nakatsuji2015,Kuebler2014}. Futhermore, a magnetic inverse spin Hall effect~\cite{Kimata2019} and large anomalous Nernst effect~\cite{Ikhlas2017} have been discovered in \MnSn{}. 
While in \MnGe{}, a similar chiral spin structure combined with Weyl points lying closer to the Fermi level produces an even larger AHE despite its moderate spin–orbit coupling~\cite{Nayak2016}. 
Compared to the more extensively studied \MnSn{} and \MnGe{}, which show topological behavior within a high-symmetry hexagonal structure, \MnGa{} is distinguished by a temperature-dependent magnetostructural transition (MST)~\cite{KREN1970,NIIDA1983,Boeije2017,Song2021}, i.e., a coupled crystal and magnetic transition, as shown in Fig.~\ref{fig:schematics}b,~c. 
\MnGa{} is isostructural with \MnSn{} and \MnGe{} at high temperatures and has been reported to exhibit a 120$^{\circ}$ chiral antiferromagnetic order below $T_\text{N1}\approx470$~K~\cite{KREN1970}. 
Upon cooling to the lower MST temperature $T_\text{N2}\approx$120~K~\cite{Song2021} or 170~K~\cite{NIIDA1983,Boeije2017}, \MnGa{} undergoes a structural transition, reported as either orthorhombic~\cite{NIIDA1983,Song2021} or monoclinic~\cite{Boeije2017}, accompanied by a magnetic reconfiguration that gives rise to a substantial net ferromagnetic moment~\cite{Song2021,Liu2017,NIIDA1983,Boeije2017}. 
Previous studies have revealed AHE and THE signatures~\cite{Liu2017,Song2021} on polycrystalline \MnGa{}, associated with the distinct crystalline and magnetic states in the temperature range $T_\text{N2}<T<T_\text{N1}$ and $T<T_\text{N2}$, respectively, suggesting a strong correlation between structural changes and topological properties. 
However, determining the true ground-state structure and spin configuration remains challenging due to the subtlety of simultaneous crystalline and magnetic symmetry lowering. 
These difficulties have hindered the identification of precise topological states below and above MST in \MnGa{}. 

In this work, we report that \MnGa{} exhibits an intrinsic topological Weyl phase transition between two distinct Weyl states, driven by a magnetostructural transition.
Using a combination of high-resolution neutron diffraction, magnetization measurements, with first-principles density functional theory (DFT) calculations, we have determined and validated the distinct crystal and magnetic structures in $T_\text{N2}<T<T_\text{N1}$ and $T<T_\text{N2}$ in \MnGa{}. 
The MST temperature in our \MnXGa{} ($x=$~2.94(1)) (hereafter referred to as \MnGa{} for simplicity), was enhanced to near room temperature ($T_\text{N2}\approx$~295~K). 
Furthermore, we find that two distinct crystalline and magnetic ordered phases can each host Weyl nodes. More importantly, the MST fundamentally alters both the multiplicity and distribution of Weyl nodes, thereby driving a topological Weyl phase transition to a distinct Weyl state accompanied by a modification of the anomalous Hall conductivity (AHC) with the emergence of non-zero $\sigma_{yz}$ component near room temperature. 
Transport measurements reveal a sign reversal of the anomalous Hall conductivity alongside a significant enhancement in its magnitude below the MST, providing experimental proof on the redistribution of the Berry curvature and a corresponding reconstruction of the Weyl-node manifold. 
The emerged THE, coexisting with AHE below a lower temperature $\approx 150$~K, highlights a wonderful unification of the real-space Berry phase arising from spin chirality and $k$-space Berry curvature inherent to the band topology. 
Our results thus demonstrate that the MST provides a compelling pathway for inducing a topological Weyl phase transition in correlated magnetic materials.

\section{Results}
\subsection{Chiral Antiferromagnetic Ordering at $T_\text{N1}$}
\begin{figure}
    \centering
    \includegraphics[width=1\linewidth]{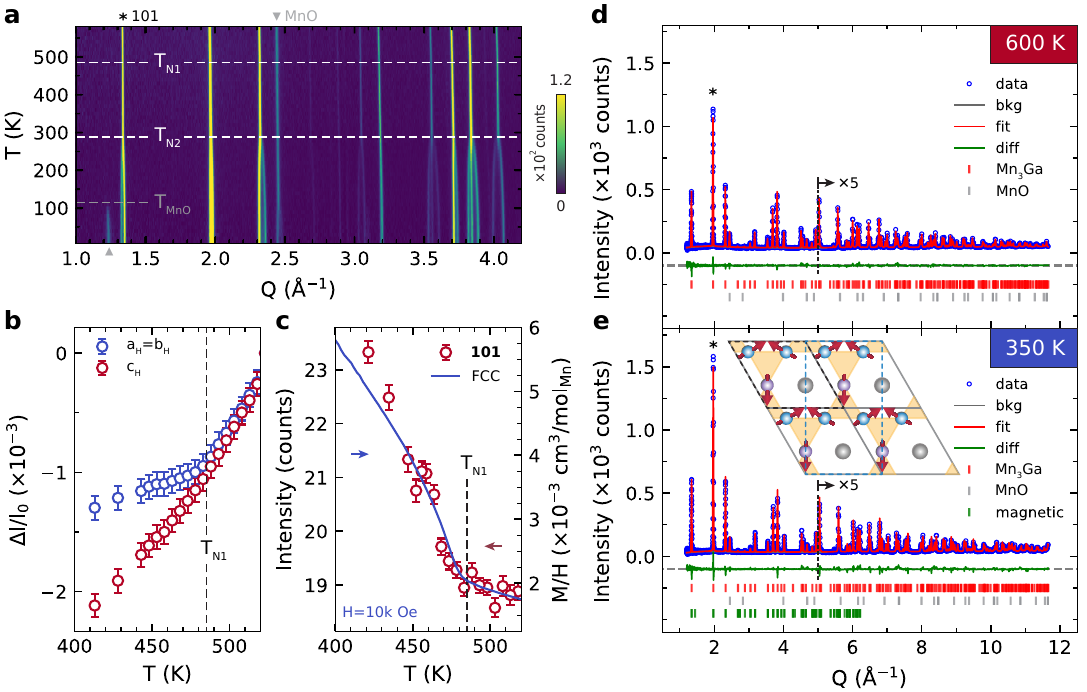}
\caption{
\textbf{Neutron diffraction and magnetization measurements of the intermediate-temperature phase ($T_\text{N2}<T<T_\text{N1}$) in \MnGa{}.
} 
\textbf{a} Neutron diffraction patterns measured from 6~K to 580~K. Black and gray asterisk symbols indicate hexagonal 101 peak and \MnO{} impurity phase. The gray arrow indicates the $\tfrac{1}{2}\tfrac{1}{2}\tfrac{1}{2}$ magnetic Bragg peak of MnO, and the temperature-dependence of  $\tfrac{1}{2}\tfrac{1}{2}\tfrac{1}{2}$ intensity is shown in Supplementary Fig.~4. \textbf{b} Temperature dependence of normalized lattice parameters. A clear lattice anomaly within the kagome plane ($a_\text{H}$ and $b_\text{H}$) is observed right below $T_\text{N1}$, indicating the onset of spin-lattice coupling. \textbf{c} Temperature dependence of field-cooled-cooling (FCC) magnetization and 101 Bragg peak intensity. \textbf{d} Rietveld refinement at 600~K confirms that the sample composition is near-stoichiometric \MnXGa{} ($x=$2.94(1)). The neutron diffraction data are well described by the hexagonal \hexsg{} structure ($Rw=$~0.0727). \textbf{e} Rietveld refinement at 350~K indicates no detectable structural phase transition. The magnetic structure is well described by the \magsgh{} magnetic space group ($Rw=$~0.0791).
}
\label{fig:TN1}
\end{figure}

We begin by investigating the magnetic structure of \MnGa{} using temperature-dependent neutron diffraction and magnetic characterization measurements (Fig.~\ref{fig:TN1}). 
As shown in Fig.~\ref{fig:TN1}a, the colormap presents the overview of phase transitions. 
At $T_\text{N1}$~=~485~K, the crystal structure retains hexagonal symmetry, but shows a reduced structural symmetry below $T_\text{N2}$~=~295~K, as indicated by the splitting of Bragg peaks. 
Lattice parameters were extracted from a series of Rietveld refinements against the diffraction data, and Fig.~\ref{fig:TN1}b shows that the lattice parameter ($a_\text{H}=b_\text{H}$) exhibits a pronounced anomaly at $T_\text{N1}$, despite the absence of any symmetry-breaking structural transition. The observed $T_\text{N1}$ is slightly higher than those~\cite{KREN1970,NIIDA1983} ($\textless$470 K) in previous reports. 
Magnetic susceptibility measurement on a polycrystalline sample (Fig.~\ref{fig:TN1}c) shows a distinct anomaly at 485~K, indicating a magnetic transition. 
Consistently, the intensity of the hexagonal 101 Bragg peak increases gradually upon cooling below $T_\text{N1}$, indicating the feature of a second-order magnetic phase transition, as shown in Fig.~\ref{fig:TN1}c. 
These results collectively indicate strong spin-lattice coupling at $T_\text{N1}$, even in the absence of a crystallographic symmetry change.

Rietveld refinements against powder neutron diffraction patterns measured above and below $T_\text{N1}$ are presented in Fig.~\ref{fig:TN1}d,~e. 
For temperatures above $T_\text{N1}$, the refinements indicate the presence of approximately 5.5 wt\% MnO impurity, which was held constant in all subsequent refinements. 
The nominal \MnGa{} phase adopts a hexagonal crystal structure with space group \hexsg{}, and the refined chemical composition yields \MnXGa{} ($x = 2.94(1)$), indicating a 2.3(3)\% of manganese deficiency relative to ideal stoichiometry. 
Upon cooling into the intermediate temperature range, $T_\text{N2}<T<T_\text{N1}$, the symmetry analysis and refinements confirm the \magsgh{} magnetic structure with magnetic propagation vector \textbf{\textit{k}}~$=$~0. 
The refined magnetic structure shows noncollinear spins confined to the kagome planes, resulting in a nearly zero net magnetic moment per unit cell (not symmetry-constrained). 
In each triangular unit, the three Mn moments are oriented $\perp \mathbf{a_\text{H}}$, $\perp \mathbf{b_\text{H}}$, and $\parallel (\mathbf{b_\text{H}} - \mathbf{a_\text{H}})$, forming a 120$^{\circ}$ chiral antiferromagnetic order as shown in the insert of Fig.~\ref{fig:TN1}e and Fig.~\ref{fig:schematics}b. 
The ordered moment per Mn site is refined to be 2.06(25)~\(\mu_\mathrm{B}\) at 350~K. 
The refined atomic positions and magnetic moments are listed in Table~\ref{tab:MagStrTN1}.

This magnetic structure differs from the 120$^{\circ}$ antiferromagnetic order previously reported in \MnGa{} where the moment direction is parallel to the crystalline axes~\cite{Boeije2017}. 
Refinements using \magsgh{} magnetic model provide improved agreement with our neutron diffraction data compared to the previously reported \magsghrpt{} magnetic model based on the early neutron powder diffraction study~\cite{Boeije2017}. The comparison is shown in Supplementary Fig.~2. 
It is worthwhile pointing out that to determine (\textbf{\textit{k}}~$=$~0) magnetic order, the high-resolution neutron diffraction data with wide $Q$ coverage is essential to obtain the reliable structural parameters, such as the scale factor, atomic positions, thermal parameters, occupancy, etc., for separating the nuclear and magnetic contributions to the same peak and distinguishing different magnetic models~\cite{Zhang2023}. 
Our DFT calculations reveal that the energy of $Cm'cm'$ configuration is $\mathrm{-30.002~eV\,f.u.^{-1}}$, which is lower than $\mathrm{-29.996~eV\,f.u.^{-1}}$ calculated for the previously proposed $P6_{3'}/m'mc'$ configuration. 
This energy difference further validates $Cm'cm'$ magnetic structure. 
As shown in Fig.~\ref{fig:HallExp}b below, we observe a small spontaneous magnetization of $0.008~\mu_B$ per Mn and a coercive field of $\approx 1$~kOe at $350$~K. 
This indicates the presence of a small uncompensated component to the moment in \MnGa{}, analogous to the small in-plane net moment along the $a_\text{H}$ axis in \MnSn{}~\cite{Nakatsuji2015} and \MnGe{}~\cite{Nayak2016}. 
Note that this minuscule moment is undetectable by neutron diffraction yet is allowed by the magnetic symmetry $Cm'cm'$ (see Table~\ref{tab:MagStrTN1}). 
Consequently, the magnetic order in the $T_\text{N2} < T < T_\text{N1}$ range is described as a nearly $120^\circ$ triangular AFM structure ($Cm'cm'$) with a slight canting that produces a weak net moment along the $a_\text{H}$ axis. 
This configuration is identical to that of iso-structural \MnGe{}~\cite{Yamada1988,Sukhanov2018} and aligns with previous DFT reports~\cite{Zhang2017}.

\subsection{Magnetostructural Phase Transition at $T_\text{N2}$}
\begin{figure}
    \centering
    \includegraphics[width=0.5\linewidth]{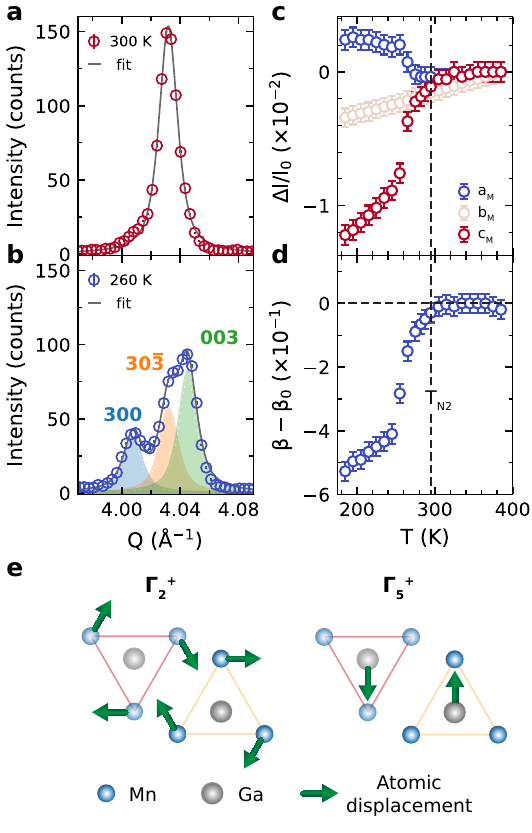}
\caption{\textbf{Structural phase transition at $T_\text{N2}$ in \MnGa{}.} \textbf{a} and \textbf{b} show hexagonal 300 Bragg peak and pseudo-Voigt function fits at 300~K and 260~K, respectively. The emergence of three distinct peaks below $T_\text{N2}$ rules out an orthorhombic distortion and confirms a hexagonal-to-monoclinic structural transition. For the fit at 260~K, the full width at half maximum (FWHM) and mixing parameter ($\eta$) of the pseudo-Voigt function were fixed to the values obtained from the 300~K fit. 
\textbf{c}--\textbf{d} Temperature dependence of normalized refined lattice parameters and monoclinic cell angle $\beta$, extracted from Rietveld refinements using the monoclinic \monosg{} space group against the neutron diffraction patterns. 
The results clearly show deviations from the high-temperature hexagonal symmetry below $T_\text{N2}$. 
The normalized lattice parameters are defined as $\Delta l/l_{0} = [l(T)-l(T_{0})]/l(T_{0})$ ($l=a_\text{M},~b_\text{M},~c_\text{M}$) to show the clear trend across $T_\text{N2}$. 
\textbf{e} Two distortion modes (\GMT{} and \GMF{}) drive the structural transition from hexagonal \hexsg{} to monoclinic \monosg{}.}
\label{fig:TN2_str}
\end{figure}

As shown in Fig.~\ref{fig:TN1}a, upon further cooling, high-resolution neutron diffraction reveals a clear splitting of Bragg peaks below $T_\text{N2}=$~295~K, indicating a symmetry-lowering distortion. 
High resolution neutron diffraction is critical to distinguish between the previously reported orthorhombic~\cite{NIIDA1983,Song2021} or monoclinic~\cite{Boeije2017} models for the low temperature structure. 
In the case of orthorhombic distortion, the hexagonal 300 peak is expected to split into two reflections (orthorhombic 330 and 060). 
In contrast, monoclinic symmetry results in three distinct peaks (monoclinic 300, 30$\bar{3}$ and 003) due to its lower symmetry and additional peak splitting. 
As shown in Fig.~\ref{fig:TN2_str}b, the observed peak splitting to three peaks at 260~K confirms the formation of a monoclinic unit cell below $T_\text{N2}$.
It is worth noting that the reflection from the \MnO{} impurity phase near Q~$=$~4~\AA$^{-1}$ is negligible and is expected to be less than 1\% of the intensity of the hexagonal 300 reflection from \MnGa{}. 
We performed pseudo-Voigt function fits on the hexagonal 300 Bragg peak above and below $T_\text{N2}$ (Fig.~\ref{fig:TN2_str}a,~b). The full width at half maximum (FWHM) and mixing parameter ($\eta$) obtained from the fit at 300~K are used as constraints for fitting the 260~K data. 
Fig.~\ref{fig:TN2_str}c,~d show the temperature dependence of lattice parameters. Clear splitting of the in-plane lattice parameters ($a_\text{M}$ and $c_\text{M}$) is observed, accompanied by a deviation of $\beta$ from 120$^{\circ}$. The Rietveld analysis on the neutron diffraction data below $T_\text{N2}$ reveals a space group {\monosg}, which shows a better refinement quality than the orthorhombic model \orthosg{} (see Supplementary Fig.~3). 
Symmetry analysis, based on the refined crystal structures and structural relationships determined using the \textsc{Bilbao Crystallographic Server}~\cite{Aroyo2006-1,Aroyo2006-2,BilbaoMaxMag}, reveals that two distortion modes, \GMT{} and \GMF{}, contribute to this monoclinic distortion. 
These modes involve atomic displacements of both Mn and Ga restricted in the Kagome plane, as shown in Fig.~\ref{fig:TN2_str}e.

It is worth noting that our \MnGa{} sample shows a substantially higher MST temperature $T_\text{N2}$ (295~K) compared with the 120~K~\cite{Song2021} and 170~K~\cite{NIIDA1983,Boeije2017} for samples with 70–74~at\% Mn, and this value is slightly higher than the transition temperature of $\approx$ 285~K reported for Mn$_{2.95}$Ga with the 74.7~at\% Mn in recent study~\cite{Boeije2017}. 
This enhancement is likely attributed to the combined effects of improved crystallinity, variations in synthesis methods and  stoichiometry of our sample. Indeed, as shown in Supplementary Fig.~1, changing the processing conditions of the \MnGa{} ingot used in this study can increase $T_\text{N2}$ to near 320~K.

\begin{figure}
    \centering
    \includegraphics[width=0.6\linewidth]{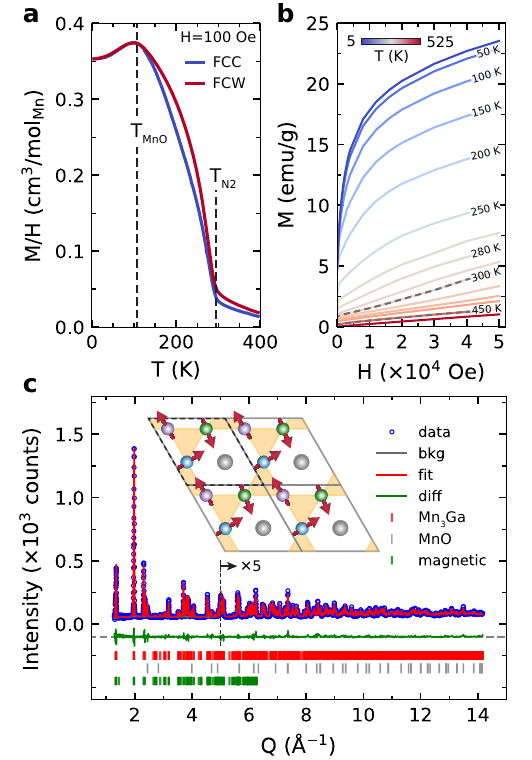}
\caption{\textbf{Magnetization and neutron diffraction results for $T<T_\text{N2}$ in \MnGa{}.} \textbf{a} Temperature-dependent magnetization measurements after field-cooled-cooling (FCC) and field-cooled-warming (FCW) protocols with $H=$100~Oe shows the onset of magnetic transition near $T_\text{N2}=$~295~K. The magnetization drops slightly near 110~K due to the magnetic transition of \MnO{} impurity phase. \textbf{b} Temperature-dependent $M(H)$ curves measured from 5 to 525~K. \textbf{c} Rietveld refinement for neutron diffraction result at 30~K ($Rw=$~0.0769) using the \magsgl{} magnetic space group.}
\label{fig:TN2}
\end{figure}

Concurrent with this structural distortion, a new magnetic ground state emerges at $T_\text{N2}$, as indicated by the onset of increasing magnetization upon cooling as shown in Fig.~\ref{fig:TN2}a and changes of the magnetic peak intensities (see Fig.~\ref{fig:TN1}a). The rapid increase of the magnetization and the pronounced  deviation from linear $M(H)$ behavior below $T_\text{N2}$ in Fig.~\ref{fig:TN2}b indicate the emergence of a ferromagnetic component below $T_\text{N2}$. 
Our neutron diffraction results shows that the propagation vector remains zero (\textbf{\textit{k}}~$=$~0). 
We found neutron powder diffraction alone cannot unambiguously determine the magnetic structure due to overlapping nuclear and magnetic peaks and the subtlety of the simultaneous lowering of crystalline and magnetic symmetry. 
Only two magnetic models (A and H), shown in Supplementary Fig.~7, yielded excellent yet comparable refinement quality. 
To identify the correct magnetic structure, we combined the refinement results with our first-principles calculations including spin-orbit coupling (DFT+SOC). 
The DFT results validated magnetic model A since it exhibits the lowest energy among nine different magnetic orders in the monoclinic phase (see more details in Supplementary information and Supplementary Figs.~6 and 7). 
The Rietveld refinement against the neutron diffraction data at 30~K using the magnetic model A (\magsgl{}) is presented in Fig.~\ref{fig:TN2}c. 
The fitted atomic positions and magnetic moments are listed in Table~\ref{tab:MagStrTN2}. 
As shown in the inset of Fig.~\ref{fig:TN2}c and Fig.~\ref{fig:schematics}c, it is a canted AFM structure with substantial net moment ($\approx$ 0.71~$\mu_\text{B}$ per Mn). 
The moments of Mn(1) and Mn(2) nearly cancel each other, and the net moment arises primarily from Mn(3), oriented nearly perpendicular to $\mathbf{a_\text{M}}$ in the monoclinic $a_\text{M}c_\text{M}$ plane, which elucidates the origin of the ferromagnetic component observed in the magnetization. 
The ordered moment per Mn site at 30~K is refined to be $\approx$ 2.56~$\mu_\text{B}$.
Note that this magnetic structure differs significantly from that proposed in the previous report~\cite{Boeije2017}. 
In addition, the decrease of the magnetization in Fig.~\ref{fig:TN2}a below $\approx$~110~K is associated with the AFM transition of impurity phase MnO, as evidenced by the the appearance of AFM peak $\frac{1}{2}\frac{1}{2}\frac{1}{2}$ of MnO below its N\'eel temperature $T_\text{MnO}=$~110~K (see Fig.~\ref{fig:TN1}a and Supplementary Fig.~4), consistent with previous reports~\cite{Roth1958,Pomjakushin2024}. 

\subsection{Anomalous Hall effects and Weyl states above/below MST}
\begin{figure}
    \centering
    \includegraphics[width=0.6\linewidth]{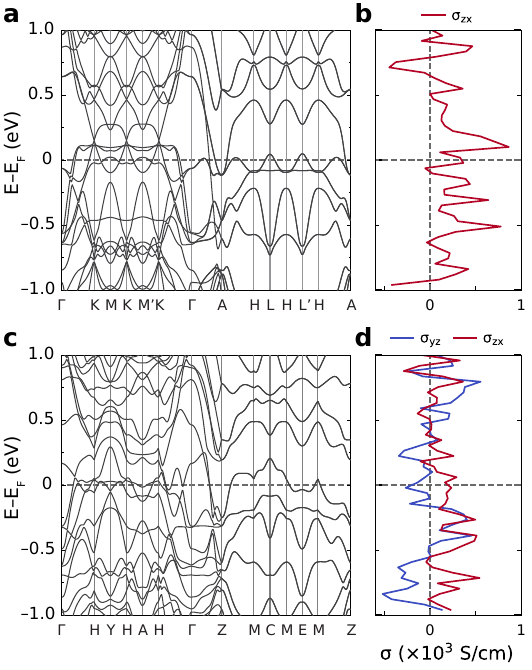}
\caption{
\textbf{Electronic band structure and AHC from DFT calculations of \MnGa{}.} 
\textbf{a},~\textbf{b} and \textbf{c},~\textbf{d} are for the hexagonal phase and the monoclinic phase, respectively. 
\textbf{a} and \textbf{c} show the band dispersions. 
We used the SeeK-path tool \cite{Hinuma2017} to determine high-symmetry ${\rm \bf k}$ points, 
which are also indicated in Fig.~\ref{fig:WeylNodes}\textbf{a} and \textbf{c}. 
\textbf{b} and \textbf{d} show the numerical results of AHC in the hexagonal phase and the monoclinic phase, respectively. 
Because of the high symmetry, only $\sigma_{zx}$ is nonzero in the hexagonal phase as shown in \textbf{b}. 
In the monoclinic phase, there are obvious changes of AHC, with the emergence of nonzero  $\sigma_{yz}$ in \textbf{d}.
}
\label{fig:HEXvsMONO}
\end{figure}

With the established crystal and magnetic symmetries,
we carried out first-principles calculations to locate potential Weyl nodes, determine their symmetry characteristics, and evaluate the associated anomalous Hall effect (AHE). 
The electronic band structure in the intermediate-temperature phase ($T_\text{N2}<T<T_\text{N1}$) is shown in Fig.~\ref{fig:HEXvsMONO}a. 
Since \MnGa{} shares the same symmetry as \MnSn{} and \MnGe{} above $T_\text{N2}$, their band structures are correspondingly similar. 
However, reflecting the smaller number of total electrons, the Fermi level in \MnGa{} is lower by $\sim 0.4$~eV. 
This result is fully consistent with that in Ref.~\cite{Zhang2017} (Fig.~3(a)). 
Because of the spatial inversion symmetry ($\cal I$) as well as the combined symmetry 
involving the mirror reflection with respect to a kagome plane and the time reversal (${\cal M}_z {\cal T}$), 
all bands on the $k_z = 2\pi/c$ plane have twofold degeneracy (see A–H–L–H–L'–H–A line in Fig.~\ref{fig:HEXvsMONO}a). 
The anomalous Hall conductivity for the hexagonal phase is shown in Fig.~\ref{fig:HEXvsMONO}b. Mirror reflection symmetry about the $y$ axis (${\cal M}_y$) forbids both $\sigma_{xy}$ and $\sigma_{yz}$, allowing only $\sigma_{zx}$. 
The current $\sigma_{zx}$ in the hexagonal phase is slightly different from the previous theoretical result~\cite{Zhang2017} due to minor variations in the lattice parameters, while the overall range remains consistent. 
The $\cal I$ and ${\cal M}_z {\cal T}$ symmetries also exist in the low-temperature monoclinic phase,
leading to the twofold degeneracy in the dispersion relation on the $k_z=2\pi/c$ plane as shown in
Fig.~\ref{fig:HEXvsMONO}c (see Z–M–C–M–E–M–Z line). 
The corresponding AHC is shown in Fig.~\ref{fig:HEXvsMONO}d. 
Although the ${\cal M}_y$ is broken in monoclinic symmetry, the combined symmetry 
${\cal M}_z {\cal T}$ still remains, which prohibits only $\sigma_{xy}$, allowing  $\sigma_{zx}$ and an emergence of non-zero $\sigma_{yz}$. 

Before going into the detailed analysis of Weyl points, 
we summarize the symmetry of Weyl points dictated by the magnetic and lattice symmetries of the system. 
For \MnSn{} and \MnGe{} with 120$^{\circ}$ ordering with the net magnetic moment along the $y$ direction, 
there exist four important symmetries, 
a mirror reflection with respect to the (010) plane ${\cal M}_y$ and a half-lattice translation ($c/2$), 
a mirror reflection with respect to the (100) plane combined with the time reversal ${\cal M}_x {\cal T}$, 
a mirror reflection with respect to the (001) plane combined with the time reversal ${\cal M}_z {\cal T}$, 
as well the spatial inversion $\cal I$. 
When there is a Weyl point at $(k_x,k_y,k_z)$ with the chirality $\chi$, 
because of ${\cal M}_y$, ${\cal M}_x {\cal T}$, and ${\cal M}_z {\cal T}$, 
there exist other Weyl points at $(k_x,-k_y,k_z)$, $(k_x,-k_y,-k_z)$, and $(-k_x,-k_y,k_z)$ with the opposite chirality $-\chi$. 
Furthermore, $\cal I$ guarantees the existence of four additional Weyl points 
at $(-k_x,-k_y,-k_z)$ with the chirality $-\chi$, as well as at  $(-k_x,k_y,-k_z)$, $(-k_x,k_y,k_z)$, and $(k_x,k_y,-k_z)$ with the chirality $\chi$. 
As shown in Fig.~\ref{fig:WeylNodes}a, Weyl points in the high-temperature hexagonal phase of \MnGa{} based on our calculations clearly follow this rule. 

In the low-temperature monoclinic phase of \MnGa{}, however, ${\cal M}_y$ and ${\cal M}_x {\cal T}$ symmetries are absent, 
and only ${\cal M}_z {\cal T}$ and $\cal I$ symmetries remain. 
As a consequence of this lower symmetry, Weyl points appear as quartets at $(k_x,k_y,k_z)$ and $(k_x,k_y,-k_z)$ with the chirality $\chi$ and 
$(-k_x,-k_y,-k_z)$ and $(-k_x,-k_y,k_z)$ with the chirality $-\chi$, as shown in Fig.~\ref{fig:WeylNodes}c. 
Therefore, our results suggest two distinct Weyl states with different AHEs below and above MST. 

\subsection{Transport Signatures of the Topological Phase Transition}
To seek for the experimental proof supporting our theoretically predicted topological Weyl state transition, we performed field-dependent transport measurements on longitudinal resistivity $\rho_{xx}$ and total Hall resistivity $\rho_{yx}$, as well as $M(H)$ measurements across $T_\text{N2}$. 
Figure~\ref{fig:HallExp}a shows the longitudinal magnetoresistance, $MR = [\rho_{xx}(H)-\rho_{xx}(0)]/\rho_{xx}(0)$, which undergoes a distinct change in slope ($\propto d(MR)/dH$) across the MST at $T_\text{N2}$. 
At 350~K ($T_\text{N2}<T<T_\text{N1}$), $d(MR)/dH$ is positive and increases with magnetic fields in the 0–50~kOe region. 
However, upon cooling to 300~K near $T_\text{N2}$, the positive $d(MR)/dH$ becomes pronounced at low fields but decreases at higher fields. 
At lower temperatures, the positive $d(MR)/dH$ evolves to negative sign and MR changes to negative over the whole measurement range. This sign change in MR was not observed in previously reported polycrystalline \MnGa{} samples~\cite{Liu2017,Song2021}.
Both the positive and negative MR from 350~K down to 5~K display nearly linear behavior without saturation, which is typical in Weyl semimetals~\cite{Takiguchi2020,Ali2014,Huang2015}. 

The temperature dependence of the magnetization $M(H)$ curves is presented in Fig.~\ref{fig:HallExp}b. 
In the $T_{\text{N2}} < T < T_{\text{N1}}$ regime, the magnetization exhibits a weak magnetic hysteresis loop at low fields ($<5$~kOe). 
At higher fields, $M(H)$ shows a linear dependence on $H$ and reaches a magnitude of only $\approx0.03~\mu_{\text{B}}$ per Mn at 50~kOe (350~K), suggesting that the nearly $120^\circ$ chiral AFM order remains robust against the external field. 
Below $T_{\text{N2}}$, the spontaneous magnetization increases sharply at low fields, indicating a significant net ferromagnetic component within the distorted AFM phase. 
While the magnetization continues to rise at higher fields, it only reaches approximately $0.25~\mu_{\text{B}}$ per Mn at 50~kOe (5~K) which is an order of magnitude lower than the ordered moment of $2.56~\mu_{\text{B}}$ per Mn determined by neutron diffraction. 
This discrepancy indicates that the applied field of 50~kOe is insufficient to fully polarize the distorted AFM order into a collinear ferromagnetic state.

The temperature dependence of the total Hall resistivity is shown in Fig.~\ref{fig:HallExp}c. 
The total Hall resistivity ~\cite{Nakatsuji2015,Kiyohara2016} is composed of the ordinary Hall resistivity ($\rho_{yx}^{\text{N}}$) and the anomalous Hall effect ($\rho_{yx}^{\text{A}}$), such that:
\begin{equation}
\rho_{yx} = \rho_{yx}^{\text{N}} + \rho_{yx}^{\text{A}}=R_{0}H+S_{\text{A}}\rho_{xx}^{2}M      
\end{equation}
where $R_{0}$ is the ordinary Hall coefficient, $S_{\text{A}}$ is a field-independent parameter, $S_{\text{A}}\rho_{xx}^{2}$ is the conventional AHE coefficient proportional to magnetization. 
Note that the intrinsic $\rho_{yx}^{\text{A}}$ reflects the integral of the Berry curvature over the entire Brillouin Zone for all occupied states.

At 350~K ($T_\text{N2}<T<T_\text{N1}$), the AHE exhibits a negative (positive) sign for positive (negative) fields. 
A prominent sign reversal, accompanied by a distinct low-field hysteresis loop, is observed in $\rho_{yx}$. As the field increases, the magnitude of the Hall signal decreases linearly. 
As further illustrated in Fig.~\ref{fig:HallExp2}a, the low-field nonlinearity and sign reversal of $\rho_{yx}$ $vs$ $M$ are attributed to an additional, unconventional AHE component, $\rho_{yx}^{\text{AF}}$, which is intrinsically coupled to the nearly $120^\circ$ triangular AFM order. 
The $R_{0}$ and $S_{\text{A}}$ coefficients are determined using the method described in previous studies on \MnSn{}~\cite{Nakatsuji2015} and \MnGe{}~\cite{Kiyohara2016}. 
We then isolate the individual  $S_{\text{A}}\rho_{xx}^{2}M$, $\rho_{yx}^{\text{AF}}$, and $\rho_{yx}^{\text{A}}=S_{\text{A}}\rho_{xx}^{2}M+\rho_{yx}^{\text{AF}}$ components as a function of $M$ and $H$, as shown in Fig.~\ref{fig:HallExp2}a and b, respectively. 
While the conventional AHE component $S_{\text{A}}\rho_{xx}^{2}M$ is weak due to very low magnetization at 350~K and follows the standard proportionality to $M$, the much stronger $\rho_{yx}^{\text{AF}}$ dominants $\rho_{yx}^{\text{A}}$ and remains nearly independent with $M$ or $H$ in high field regions. 
Note that the $\rho_{yx}$ profile of \MnGa{} at 350~K bears some similarity to the averaged Hall response based on the single crystal results on the isostructural \MnSn{}~\cite{Nakatsuji2015} and \MnGe{}~\cite{Nayak2016,Kiyohara2016}, and that on polycrystalline \MnGa{}~\cite{Liu2017}.

The magnetic structure in this region involving a weak ferromagnetic component is characterized by the orthorhombic magnetic space group $Cm'cm'$ in Fig.~\ref{fig:schematics}b, which explicitly breaks time-reversal symmetry, a fundamental requirement for the Weyl points and AHE. 
By using the the hexagonal structure and $Cm'cm'$ magnetic configuration, our DFT calculations have showed a robust AHE with $\sigma_{yz}$ tensor driven by a significant enhancement of the Berry curvature localized around Weyl nodes near the Fermi level. 
The observed $\rho_{yx}^{\text{A}}$, encompassing both conventional $S_{\text{A}}\rho_{xx}^{2}M$ and unconventional $\rho_{yx}^{\text{AF}}$, provides experimental evidence to support our theoretical predictions. 

Upon cooling below $T_{\text{N2}}$, $\rho_{yx}$ shows pronounced change in the nonlinearity, indicating a distinct shift in the AHE ($\rho_{yx}^{\text{A}}$) contribution that deviates from the linear field dependence characteristic of the $\rho_{yx}^{\text{N}}$. 
This occurs near room temperature in our nearly stoichiometric \MnGa{}, significantly higher than the approximately 140~K observed in previous report on \MnGa{}~\cite{Liu2017}. 
In addition, in sharp contrast to the high-temperature region, the $\rho_{yx}$ becomes positive (negative) under positive (negative) fields at low temperatures, indicating a global sign reversal of the $\rho_{yx}^{\text{A}}$. 
Notably, a pronounced hump-like feature emerges at moderate fields ($<10$~kOe) below $\sim$ 150~K and vanishes at higher fields (see Fig.~\ref{fig:HallExp2}d and Supplementary Fig.~5c)—both of which are characteristic hallmark of the topological Hall effect (THE)~\cite{Gerber2018,Kimbell2022,Li2013}. 
Although THE was previously reported in polycrystalline \MnGa{}~\cite{Liu2017}, it was associated with a different orthorhombic structure and occurred at a lower temperature, around 100~K.

Since the high-field region can be well described by $R_{0}H+S_{\text{A}}\rho_{xx}^2 M$, we determine the coefficients $R_{0}$ and $S_{\text{A}}$ by fitting the high-field data to the linear relation $\rho_{yx}/H = R_{0} + S_{\text{A}} (\rho_{xx}^2 M / H)$~\cite{Liu2017} (see Supplementary Fig.~5d). 
The subtracted component $\rho_{yx}-R_{0}H-S_{\text{A}}\rho_{xx}^{2}M$ is shown in Fig.~\ref{fig:HallExp2}c-d, which is predominantly dominated by THE showing a clear hump at low fields that vanishes at high fields. 
Consequently, the Hall resistivity at low temperatures is modeled as $\rho_{yx} = R_{0}H + \rho_{yx}^{\text{A}} + \rho_{yx}^{\text{T}} =R_{0}H + S_{\text{A}}\rho_{xx}^2 M + \rho_{yx}^{\text{T}}$ where the $\rho_{yx}^{\text{A}}$ is well-described by the conventional AHE mechanism ($S_{\text{A}}\rho_{xx}^2 M$). 
As shown in Fig.~\ref{fig:HallExp2}c-d, there is a dramatic increase of the $S_{\text{A}}\rho_{xx}^2 M$ component at 100 K compared to 350 K, which correlates with a significantly increased net moment. In addition to the emergence of a distinct $\rho_{yx}^{\text{T}}$, the AHE ($\rho_{yx}^{\text{A}}$) at 100~K exhibits an opposite sign and different magnitudes from those at 350~K (see Fig.~\ref{fig:HallExp2}a-b).

To further demonstrate the topological contribution to the Hall response, we track the temperature evolution of the Hall conductivity, $\sigma_{xy}= -\rho_{yx}/\rho_{xx}^{2}$ ($\rho_{xx}>>\rho_{yx}$, see Supplementary Fig.~5a). 
As shown in Fig.~\ref{fig:HallExp}d, $\sigma_{xy}$, which represents the average anomalous Hall conductivity across the three tensor components of the theoretical AHC, shows a significant change in nonlinearity and a sign reversal below $T_\text{N2}$, attributed to the topological Weyl transition. 
In comparison to the maximum $\sigma_{xy}^{A}$ magnitude of 13.7~($\Omega$cm)$^{-1}$ observed at 350~K (above $T_\text{N2}$), the AHC increased significantly to 25.8~($\Omega$cm)$^{-1}$ at 100~K (below $T_\text{N2}$), as shown in Supplementary Fig.~5b. 
This behavior is consistent with the calculated enhancement of the AHC resulting from the appearance of extra $\sigma_{yz}$ tensor (Fig.~\ref{fig:HEXvsMONO}d), which is originated from the simultaneous structural and magnetic structural transformation at $T_\text{N2}$. 
Therefore, our results on $\rho_{yx}$ and $\sigma_{xy}$ provides transport evidence that the MST reorganizes the Berry-curvature distribution and reconstructs the Weyl-node manifold in \MnGa{}.

\section{Discussion}
\subsection{Unique Mechanism of MST Phase Transition in Mn$_{3}$X family}
To gain deeper insight into the mechanisms driving phase transition in \MnGa{}, it is instructive to compare physical properties among those related compounds in the same family. 
All three materials \MnX{}, can be stabilized in the hexagonal $D0_{19}$ structure (Fig.~\ref{fig:schematics}a) and their magnetic ordering breaks time-reversal symmetry~\cite{Sticht1989,Brown1990,Nayak2016,Liu2017,Boeije2017}. 
Both theoretical and experimental studies have shown that the chiral antiferromagnetic order in \MnSn{}, \MnGe{}, and \MnGa{} gives rise to a pronounced anomalous Hall effect (AHE) and spin Hall effect (SHE), driven by Berry curvature associated with their magnetic order~\cite{Zhang2017}. 
However, \MnGa{} is distinguished from \MnSn{} and \MnGe{} by its electronic configuration, possessing one fewer valence electron, which leads to two key consequences.

First, while the electronic band structures of \MnX{} are qualitatively similar, differences in valence electron count significantly impact their topological behavior. 
In \MnGe{}, the Fermi level resides close to several Weyl nodes, resulting in strong Berry curvature and a large intrinsic AHE that persists up to high temperatures~\cite{Nayak2016}. 
\MnSn{} exhibits stronger spin–orbit coupling, which leads to partial annihilation and displacements of Weyl nodes slightly away from the Fermi level~\cite{Nakatsuji2015}. 
By contrast, the reduced electron count in \MnGa{} lowered the Fermi energy, suppressing the net Berry curvature at the Fermi surface and weakening the AHE, while potentially enhancing the spin Hall conductivity through redistribution of the Berry curvature proposed by previous study~\cite{Zhang2017}. 

Second, while the hexagonal structure in \MnGe{} and \MnSn{} persists down to the lowest temperature, the presence of a magnetostructural phase transition in \MnGa{} suggests that orbital degeneracy is lifted~\cite{Liu2017,Boeije2017}. 
This indicates a strong coupling between lattice, orbital and spin degrees of freedom in \MnGa{}, which is absent in \MnGe{} and \MnSn{}. 
What is the driving force for the magnetostructural phase transition at $T_\text{N2}$ in \MnGa{}? 
According to our DFT calculation, the total energy of the most stable magnetic structure in the monoclinic phase is
$\mathrm{-30.043~eV\,f.u.^{-1}}$, while that of the 120$^{\circ}$ triangular AFM structure in the hexagonal phase is $\mathrm{-30.002~eV\,f.u.^{-1}}$ (both are computed at $T=0$). 
The magnetostructural phase transition between the two phases might be reasonably accounted for thermal effects, including the reduction in the ordered moment and phonon excitations, that could overcome the small energy difference $\mathrm{0.041~eV\sim475}$~K. 
The unique combination of monoclinic structure and the canted AFM order with net ferromagnetic component in \MnGa{} leads to a redistribution of Weyl nodes and drives a topological Weyl phase transition, which was not observed in \MnGe{} and \MnSn{}.

\subsection{Emergent anomalous phase at low temperatures and fields}
The emergence of the THE is rooted in the scalar spin chirality of non-coplanar magnetic textures and acts as a source of real-space Berry phase.\cite{Kimbell2022,Li2023} 
While neutron diffraction at zero field reveals a distorted AFM order with monoclinic magnetic symmetry, this configuration remains coplanar. 
Consequently, the scalar spin chirality, defined by $\textbf{S}_{i} \cdot (\textbf{S}_{j} \times \textbf{S}_{k})$~\cite{Kimbell2022,Li2023} for a spin triad, is restricted to zero. 
The observation of a THE at low fields ($<10$~kOe) suggests that the ground-state distorted AFM order undergoes a field-induced transition to a non-coplanar state with non-zero spin chirality. 
A plausible scenario is that the applied field induces spin canting along the out-of-plane $c_\text{H}$ axis. 
Theoretical calculations have demonstrated that such out-of-plane canting can generate substantial Berry curvature along nodal lines in $k$-space, enhancing the AHE and potentially fostering the formation of new Weyl points, as seen in the hexagonal phase of \MnSn{}~\cite{Li2023}. 
Thus, while the THE is fundamentally a real-space Berry phase phenomenon, the underlying spin canting simultaneously modifies the $k$-space Berry curvature. 
The coexistence of AHE and THE in \MnGa{} therefore represents a profound convergence of real-space and momentum-space topological effects.

In contrast to the high-temperature nearly 120$^{\circ}$ triangular AFM phase, which exhibits no THE up to 50~kOe, the THE becomes prominent below $\approx 150$~K. 
This temperature corresponds to the regime where the rate of increase in magnetization begins to diminish upon cooling (Supplementary Fig.~1). 
Furthermore, the THE coincides with the rapid low-field increase in magnetization observed in $M(H)$ curves below 10~kOe (Fig.~\ref{fig:HallExp}b). 
These correlations indicate that the response of the net FM component to the external fields may be closely related the spin chirality and the appearance of THE in \MnGa{}. 
At higher fields exceeding $\approx 10$~kOe, the $M(H)$ curves exhibit a continuous increase with reduced $dM(H)/dH$, reflecting the response of the primary AFM component to the field. 
The disappearance of the THE in this regime suggests that the spin chirality reverts to zero. 
Furthermore, the THE is often considered one hallmark of the skyrmion phase~\cite{Kimbell2022}, like the A phase in MnSi~\cite{Neubauer2009}. Thus, an anomalous phase
with spin chirality exists at low fields below approximately 150 K in \MnGa{}. Our results may motivate further investigation into the field-dependent magnetic structures/spin chirality, topological states and THE-related phenomena in \MnGa{}.

\subsection{MST-Driven Topological Weyl phase transition}
\begin{figure}
    \centering
    \includegraphics[width=1\linewidth]{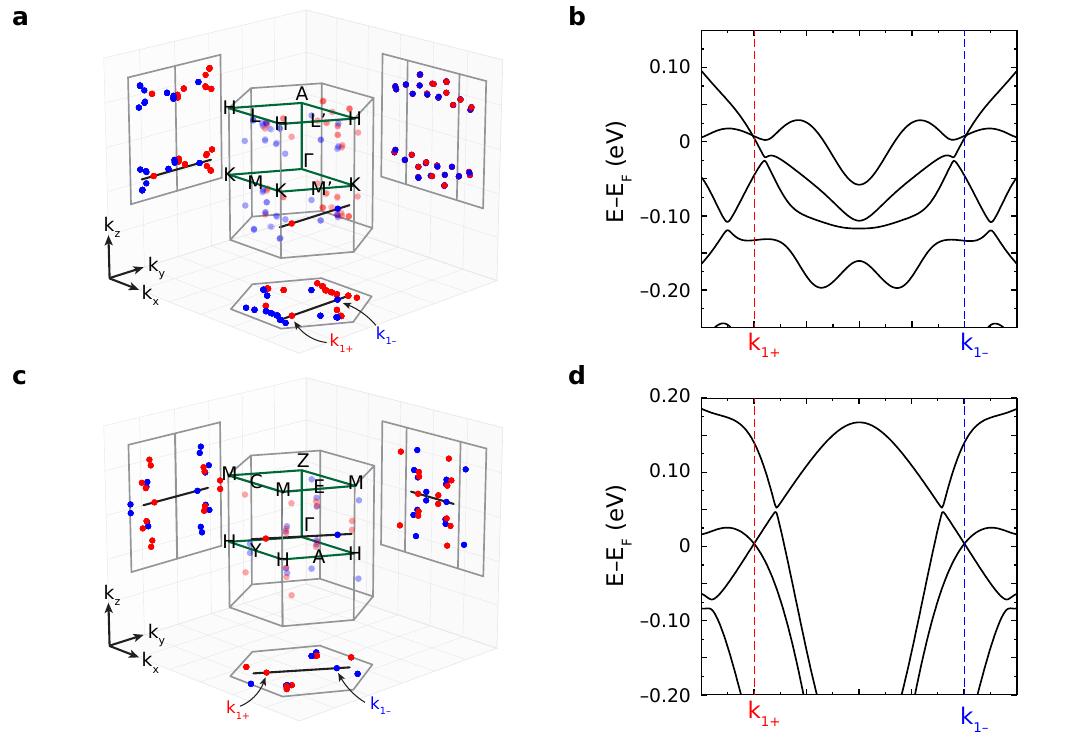}
\caption{\textbf{Weyl points and dispersions of \MnGa{}.}
\textbf{a} and \textbf{c} show Weyl points, which energy is within $\pm 0.01$~eV from $E_\text{F}$. 
\textbf{b} and \textbf{d} show examples of band dispersions along momentum cuts indicated by black lines in \textbf{a} and \textbf{c} across Weyl points at ${\rm \bf k}_{1 \pm}$, showing type-II (\textbf{b}) and type-I (\textbf{d}) features in hexagonal and monoclinic phases, respectively.}
\label{fig:WeylNodes}
\end{figure}

\begin{figure}
    \centering
    \includegraphics[width=1\linewidth]{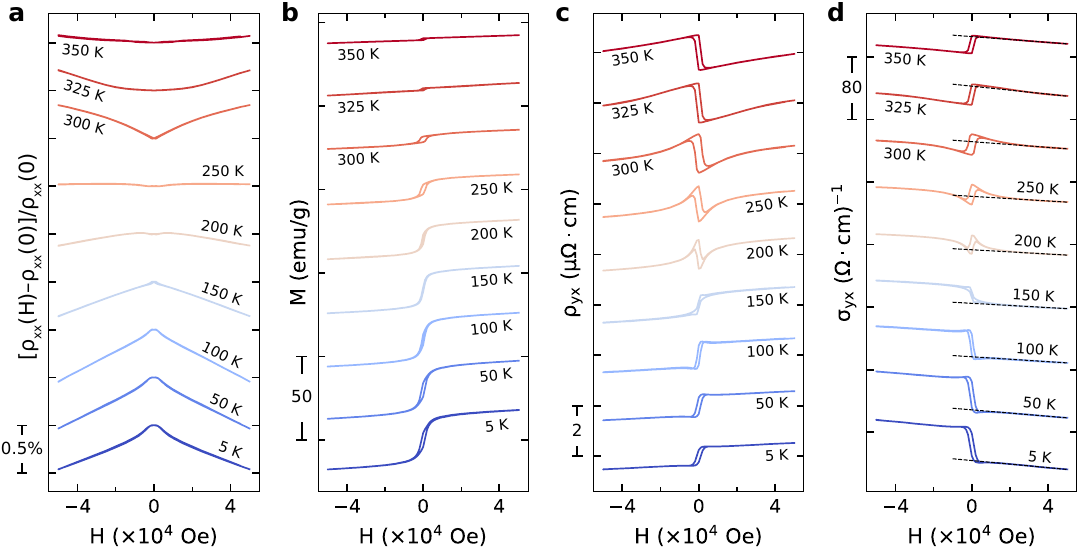}
\caption{
\textbf{Evolution of transport and magnetic properties across $T_\text{N2}$ in polycrystalline \MnGa{}.} 
\textbf{a} Field-dependent longitudinal magneto-resistance (MR). 
\textbf{b} Magnetization $M(H)$. 
\textbf{c} Hall resistivity $\rho_{yx}$. 
\textbf{d} Hall conductivity $\sigma_{xy}(H)$.
For visual clarity, curves at different temperatures are vertically offset by 0.5$\%$ in \textbf{a}, 50~emu/g in \textbf{b}, 2~$\mu \Omega \cdot$cm in \textbf{c}, and 80~$\Omega^{-1}$cm$^{-1}$ in \textbf{d}.
}
\label{fig:HallExp}
\end{figure}

\begin{figure}
    \centering
    \includegraphics[width=0.8\linewidth]{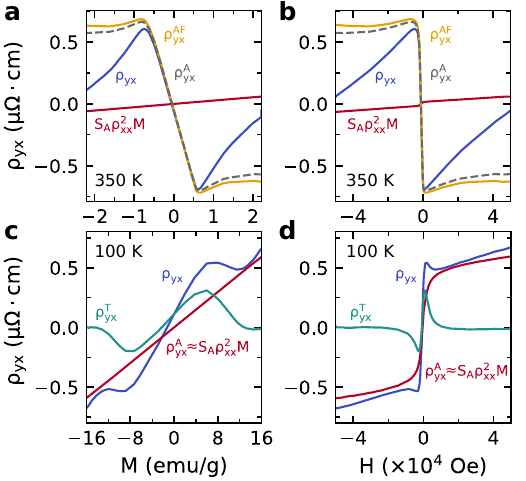}
\caption{
\textbf{Decomposition of the Hall response at 350 and 100~K.} 
Magnetization-dependent \textbf{a} and field-dependent \textbf{b} of total Hall resistivity $\rho_{yx}$, and conventional AHE components ($S_{\text{A}}\rho_{xx}M$) and antiferromagnetic ($\rho_{yx}^{\text{AF}}$) contributions at 350~K. The dashed line indicated the total AHE component ($\rho_{yx}^{\text{A}}=S_{\text{A}}\rho_{xx}M+\rho_{yx}^{\text{AF}}$). 
Magnetization-dependent \textbf{c} and field-dependent \textbf{d} of total Hall resistivity $\rho_{yx}$, total AHE components (dominated by the conventional AHE, where $\rho_{yx}^{\text{A}}\approx S_{\text{A}}\rho_{xx}M$), and topological ($\rho_{yx}^{T}$) contributions at 100~K.
}
\label{fig:HallExp2}
\end{figure}

First-principles calculations including spin-orbit coupling demonstrate that both the hexagonal and monoclinic phases of \MnGa{} host symmetry-protected Weyl nodes, though with fundamentally different distributions governed by crystal and magnetic symmetries. 
In the hexagonal phase ($T_\text{N2}<T<T_\text{N1}$), the magnetic structure we observe is nearly $120^\circ$ chiral AFM order with a weak in-plane FM component, consistent with theoretical model reported previously~\cite{Boeije2017}. 
The combination of mirror, time-reversal and inversion symmetries produces a high multiplicity of Weyl nodes of \MnGa{} arranged in symmetry-related sets, generating a Berry curvature landscape similar to that predicted for \MnSn{} and \MnGe{} with many pairs of type-II Weyl nodes identified~\cite{Yang2017,Chen2021,Soluyanov2015}. 
Figure~\ref{fig:WeylNodes}b shows examples of tilted type-II Weyl nodes in the hexagonal phase for \MnGa{}. 
Our calculations demonstrate that the integration of the Berry curvature over the entire Brillouin Zone for all occupied states, including the region around the Weyl points in this band structure gives rise to a substantial $\sigma_{zx}$ component of the AHE (Fig.~\ref{fig:HEXvsMONO}b). 
This theoretical result is corroborated by our experimental observations of the AHE, which encompass both the conventional ferromagnet-like contribution and an unconventional AHE component (Figs. \ref{fig:HallExp}c–d and \ref{fig:HallExp2}a-b).

In $T<T_\text{N2}$, our study resolves the previously debated crystal and magnetic structures. 
This newly identified spin configuration with a strongly distorted AFM order, combined with the symmetry-lowering structural distortion below $T_\text{N2}$, breaks ${\cal M}_y$ and ${\cal M}_x {\cal T}$ symmetries while preserving ${\cal M}_z {\cal T}$ and inversion ($\cal I$). 
This symmetry reduction reorganizes the Weyl node network with redistributed momentum-space positions and alters chirality arrangements, as shown in Fig.~\ref{fig:WeylNodes}. 
Notably, the symmetry-driven reconfiguration of the Weyl nodes could be accompanied by the change in the type of Weyl points from strongly tilted type-II character~\cite{Soluyanov2015} in the hexagonal phase (Fig.~\ref{fig:WeylNodes}b) to the upright type-I character~\cite{Soluyanov2015} in the monoclinic phase (Fig.~\ref{fig:WeylNodes}d), representing a rare example of a temperature-dependent intrinsic Weyl phase transition. 
This transition is characterized by a modification of the AHC tensor, specifically the emergence of an additional $\sigma_{yz}$ component alongside $\sigma_{zx}$, which leads to an enhancement of the total AHC, as shown in Fig.~\ref{fig:HEXvsMONO}d. 
Experimentally, we observed a reversal in the sign of the AHC and a twofold increase in its maximum magnitude at low temperatures, as shown in Fig.~\ref{fig:HallExp}d and Supplementary Fig.~5b, compared to the profiles observed at 350~K in the hexagonal phase. 
Our theoretical results showed that the Berry curvature distribution and the Weyl-node topology in the monoclinic phase are distinct from those in the hexagonal phase, leading to the observed changes in the anomalous Hall effects above/below $T_\text{N2}$. 
Furthermore, the exclusive appearance of the THE in the monoclinic phase suggests that the Weyl state, as well as the interplay between momentum-space and real-space Berry curvatures, can be effectively tuned by magnetic fields. 
The topological reconstruction, driven by a magnetostructural transition, demonstrates how the concurrent breaking of crystal and magnetic symmetries can reshape the electronic structure and generate novel topological Weyl state and THE.

\subsection{Engineering Magnetostructural Transitions}
Our results have demonstrated that magnetostructural phase transitions play a decisive role in governing the topological properties of \MnGa{}. 
MSTs are known to occur across a broad class of materials, such as Heusler Ni–Mn–$X$ ($X=$Ga, Sn) alloys~\cite{Ullakko1996,Krenke2005}, FeRh intermetallic~\cite{Kouvel1962,Annaorazov1996}, MnAs compound~\cite{Kato1983}, perovskite manganites~\cite{Schiffer1995} and spinel oxide MnV$_{2}$O$_{4}$~\cite{Garlea2008}. 
Interestingly, MSTs can be triggered through targeted chemical design even in cases where MST is not present in the parent compounds~\cite{Trung2010,Liu2012,Liu2016}. 
For example, the parent compound MnCoGe lacks magnetostructural coupling, exhibiting a substantial temperature separation of around 300~K between its magnetic transition temperature ($T_\text{C}$), and structural transition temperature ($T_\text{S}$). 
However, introducing only a few percent of interstitial boron can tune the magnetic and structural transitions to coincide for inducing the occurrence of MST~\cite{Trung2010}. 
In another example of the MnNiSi–CoNiGe alloy, co-substitution of Co and Ge significantly shifts the structural transition temperature, and aligns it with the magnetic ordering temperature, thereby driving the MST and enhancing magneto-responsive  performance~\cite{Liu2016}. 
Furthermore, the MSTs can be controlled and tuned effectively through chemical substitution~\cite{Liu2012,Aryal2017}, heat treatment~\cite{Czaja2016,Chen2019}, magnetic field~\cite{Pecharsky2001}, pressure~\cite{Manosa2010}, or film engineering~\cite{Daeweritz2006}.  
Therefore, investigating magnetostructural transitions opens broad possibilities for realizing topological quantum phases governed by simultaneously broken crystal and magnetic symmetries. 

In summary, we report \MnGa{} as a rare system that undergoes an intrinsic topological Weyl phase transition between two distinct Weyl states driven by a magnetostructural transition near room temperature. 
The magnetostructural transition near room temperature involves a symmetry-lowering lattice distortion and a magnetic structural transformation, leading to a Weyl phase transition from a primary type-II Weyl state to a distinct Weyl state, and a pronounced change of both signs and magnitudes of the anomalous Hall effect, as well as the emergence of topological Hall effect. 
Associated with the THE, an anomalous phase with spin chirality is identified at low fields ($\textless$10 kOe) below approximately 150~K. 
Our findings reveal a compelling route to alternate emergent topological states through intrinsic magnetostructural transition in correlated quantum materials. Further studies on \MnGa{} using angle-resolved photoemission spectroscopy, scanning tunneling microscopy, AHE measurements on single crystals, as well as field tuning and film engineering, are required to bring up further new physics and potential applications through the controlled switching and manipulation of two distinct Weyl states. 
Our study should motivate future theoretical and experimental studies on other materials in which the magnetostructural transition occurs or can be induced for identifying and realizing novel topological Weyl states. 


\section{Methods}
\subsection{Sample preparation}
High-purity manganese (99.95\%) and gallium (99.9999\%) were combined in a 3:1 molar ratio and arc-melted to form an initial alloy. 
The resulting ingot was sealed in an evacuated fused silica ampoule and annealed at 673~K for 10 days, resulting in the formation of \MnGa{} in the tetragonal phase. 
To obtain the hexagonal phase, the ingot was resealed under vacuum, heated to 873~K for 2 hours, and subsequently quenched in room-temperature water. 

Further preparation for the powder neutron diffraction sample involved grinding part of the ingot into powder using a percussion mortar and pestle, followed by ball milling in a tungsten carbide-lined crucible with tungsten carbide balls. 
The powder was annealed (873 K) for 17 hours and quenched in room-temperature water, resulting in well-crystallized hexagonal \MnGa{} powder.

To investigate the impact of ball milling on the magnetic transition temperature ($T_{\text{N2}}$), a separate bulk fragment was cut from the ingot and annealed directly at 873~K for 17~hours without prior milling. 
Magnetization measurements revealed an increased transition temperature of $T_{\text{N2}}\approx$~320~K, as shown in Supplementary Fig.~1.

Further preparation for the electrical transport measurements involved cutting and grinding part of the ingot into rectangular parallelepiped samples. 
To relieve surface strain and damage induced during the shaping process, these samples underwent an initial annealing at 873~K for 17 hours followed by a water quench. 
A subsequent annealing step was performed at 873~K for 6~hours. 
The resulting $T_{\text{N2}}$ was approximately 310~K, close to the transition temperature observed in the powder samples used for neutron diffraction.

\subsection{Neutron diffraction}
Neutron powder diffraction measurements were carried out using the high-resolution neutron diffractometer POWGEN at the Spallation Neutron Source (SNS), Oak Ridge National Laboratory. 
A 3.5~g powder sample of \MnGa{} was loaded into a vanadium can, which was subsequently mounted in a cryofurnace (JANIS) to cover temperature region from 30~K to 600~K. 
An orange cryostat was also used for complementary data collection from 2~K to 300~K. 
Isothermal neutron diffraction measurements were conducted at 30, 200, 350, and 600~K using the neutron frame 1 (center wavelength: 0.8~\AA) to cover wide $Q$ region from 1 to 14~\AA$^{-1}$. 
To specially check the small distortion, the high-resolution setup with neutron frame 2 (center wavelength: 1.5~\AA) and neutron frame 3 (center wavelength: 2.665~\AA) were also used for the measurements. 
At each measured temperature, sufficient waiting time was allowed to ensure thermal stability. 
Continuous temperature-dependent measurements from 30 to 600~K were conducted using a cryofurnace (JANIS) to capture the evolution of structural and magnetic Bragg peaks across two phase transitions.

Neutron powder diffraction data were used to refine the crystal and magnetic structures of \MnGa{} using the \textsc{GSAS-II} software package~\cite{GSAS2}. 
$|M|$ represents the size of the ordered magnetic moment vector $\textbf{M} = (m_x, m_y, m_z)$. 
It was calculated based on the $(m_x, m_y, 0)$ components for the hexagonal phase or the $(m_x, 0, m_z)$ components for the monoclinic phase, incorporating the respective inter-component angles. 
The GSAS-II software computed these values directly, which were subsequently extracted from the refinement output (\texttt{.lst}) files. 
To obtain the average net moment per Mn, the vector sum of all Mn moments in the magnetic unit cell was normalized by the six Mn atoms contained within one magnetic unit cell. 
The \textsc{Bilbao Crystallographic Server}~\cite{Aroyo2006-1,Aroyo2006-2,BilbaoMaxMag} and \textsc{Isodistort}~\cite{Campbell2006} were used to determine the structural distortion modes and symmetry-allowed magnetic space groups.

\subsection{Magnetic characterization and transport measurements}
Magnetization measurements were performed using a Magnetic Property Measurement System (MPMS-XL) from Quantum Design. 
For measurements between 5 and 395~K, 46~mg of powder sample was loaded into gelcaps and mounted into a plastic drinking straw. 
For measurements between 300 and 550~K, 30~mg of powder sample was loaded in a fused silica tube.

Electrical transport measurements were carried out in a Quantum Design Physical Property Measurement System (PPMS) and Quantum Design Dyancool using the Electrical Transport Option (ETO). 
Two samples were measured and showed consistent results. 
Magnetic field dependent measurements in the longitudinal resistivity ($\rho_{xx}$) and Hall ($\rho_{yx}$) configurations were collected under isothermal conditions by stepping the applied field from 50~kOe to –50~kOe then back to 50~kOe. 
For $\rho_{xx}$, data from a sample with leads attached in a longitudinal configuration was symmetrized by averaging the results from the decreasing-field scan and the increasing-field scan (to remove Hall contributions arising from lead misalignment). 
For $\rho_{yx}$, data from a sample with leads attached in a transverse configuration were antisymmetrized by taking one half of the difference between the resistance measured on decreasing and on increasing the field (to remove longitudinal contributions arising from lead misalignment). 
Note that in Fig.~\ref{fig:HallExp} full loops of symmetrized and antisymmetrized transport data are shown. 
Since they were determined from a single loop of raw data, the branches in the full loops of processed data are related and do not represent independent data collected on increasing and decreasing the field. 
Corresponding isothermal magnetization curves were measured on a piece cut from the \MnGa{} ingot and heat treated in the same way as the transport samples. 

\subsection{Density functional theory}
To gain insight into the magnetic ordering and potential Weyl physics in the regimes $T_\text{N2}<T<T_\text{N1}$ and $T<T_\text{N2}$, 
we carried out DFT calculations for \MnGa{} using the hexagonal and monoclinic structures obtained from neutron diffraction measurements at 350~K and 30~K, respectively. 
The projector augmented wave method \cite{Blochl1994,Kresse1999} is used with the generalized gradient approximation in 
the parametrization of Perdew, Burke, and Enzerhof \cite{Perdew1996} for exchange-correlation 
as implemented in the \textsc{Vienna {\it ab} initio simulation package} (VASP) \cite{Kresse1996a,Kresse1996b}. 
For Mn, we used a potential in which the $p$ states are treated as a valence states (Mn$_{pv}$ in the VASP distribution), 
and for Ga, a potential in which $d$ states are treated as valence states (Ga$_d$). 
For the electronic self-consistent calculations, we use a $12 \times 12 \times 12$ $\rm \bf k$-point grid and an energy cutoff of 500~eV. 
Spin-orbit coupling (SOC) is included, whereas the $+U$ correction is omitted 
because \MnGa{} is an itinerant magnetic system. To investigate the ground-state magnetic structure, we consider nine initial spin configurations that  conform to the lattice symmetry, and let them relax to stable configurations. 

Subsequently, the topological properties of \MnGa{} are investigated using an effective tight-binding model. 
We employed the \textsc{Wannier90} code \cite{Pizzi2020} to derive the maximally localized Wannier functions for both the hexagonal and monoclinic phases of \MnGa{}. 
The effective tight-binding models were refined utilizing the \textsc{WannSymm} package \cite{Zhi2022}. The locations and chiralities of the Weyl points were then analyzed using these effective tight-binding models with the \textsc{WannierTools} package \cite{Wu2018}. 
For the calculations on AHC, we used a fixed laboratory frame with $\mathbf{x} \parallel \mathbf{a_\text{H}} \parallel \mathbf{c_\text{M}}$, $\mathbf{z} \parallel \mathbf{c_\text{H}} \parallel \mathbf{b_\text{M}}$ and $\mathbf{y}$ is set by the right-hand rule (gray axes in Fig.~\ref{fig:schematics}b). 

\section{Data availability}
The datasets generated and analysed during this study are available in the Zenodo repository at https://doi.org/10.5281/zenodo.18928689. 
The repository includes neutron diffraction, magnetization, and transport data, as well as density functional theory (DFT) calculation results associated with this work. Source data underlying the figures are provided with this paper.

\section{Code availability}
The computer code supporting the findings of this study is available from the corresponding author upon reasonable request.


\clearpage 

\begin{thebibliography}{10}
\expandafter\ifx\csname url\endcsname\relax
  \def\url#1{\texttt{#1}}\fi
\expandafter\ifx\csname urlprefix\endcsname\relax\def\urlprefix{URL }\fi
\providecommand{\bibinfo}[2]{#2}
\providecommand{\eprint}[2][]{\url{#2}}

\bibitem{Weng2015}
\bibinfo{author}{Weng, H.}, \bibinfo{author}{Fang, C.}, \bibinfo{author}{Fang,
  Z.}, \bibinfo{author}{Bernevig, B.~A.} \& \bibinfo{author}{Dai, X.}
\newblock \bibinfo{title}{Weyl semimetal phase in noncentrosymmetric
  transition-metal monophosphides}.
\newblock \emph{\bibinfo{journal}{Phys. Rev. X}} \textbf{\bibinfo{volume}{5}}
  (\bibinfo{year}{2015}).

\bibitem{Wan2011}
\bibinfo{author}{Wan, X.}, \bibinfo{author}{Turner, A.~M.},
  \bibinfo{author}{Vishwanath, A.} \& \bibinfo{author}{Savrasov, S.~Y.}
\newblock \bibinfo{title}{Topological semimetal and {Fermi-arc} surface states
  in the electronic structure of pyrochlore iridates}.
\newblock \emph{\bibinfo{journal}{Phys. Rev. B}} \textbf{\bibinfo{volume}{83}},
  \bibinfo{pages}{205101} (\bibinfo{year}{2011}).

\bibitem{Xu2015}
\bibinfo{author}{Xu, S.-Y.} \emph{et~al.}
\newblock \bibinfo{title}{Discovery of a {W}eyl fermion semimetal and
  topological {F}ermi arcs}.
\newblock \emph{\bibinfo{journal}{Science}} \textbf{\bibinfo{volume}{349}},
  \bibinfo{pages}{613--617} (\bibinfo{year}{2015}).

\bibitem{Lu2015}
\bibinfo{author}{Lu, L.} \emph{et~al.}
\newblock \bibinfo{title}{Experimental observation of {Weyl} points}.
\newblock \emph{\bibinfo{journal}{Science}} \textbf{\bibinfo{volume}{349}},
  \bibinfo{pages}{622--624} (\bibinfo{year}{2015}).

\bibitem{Lv2015}
\bibinfo{author}{Lv, B.~Q.} \emph{et~al.}
\newblock \bibinfo{title}{Experimental discovery of {Weyl} semimetal {TaAs}}.
\newblock \emph{\bibinfo{journal}{Phys. Rev. X}} \textbf{\bibinfo{volume}{5}},
  \bibinfo{pages}{031013} (\bibinfo{year}{2015}).

\bibitem{Armitage2018}
\bibinfo{author}{Armitage, N.~P.}, \bibinfo{author}{Mele, E.~J.} \&
  \bibinfo{author}{Vishwanath, A.}
\newblock \bibinfo{title}{{Weyl} and {Dirac} semimetals in three-dimensional
  solids}.
\newblock \emph{\bibinfo{journal}{Rev. Mod. Phys.}}
  \textbf{\bibinfo{volume}{90}}, \bibinfo{pages}{015001}
  (\bibinfo{year}{2018}).

\bibitem{Nakatsuji2015}
\bibinfo{author}{Nakatsuji, S.}, \bibinfo{author}{Kiyohara, N.} \&
  \bibinfo{author}{Higo, T.}
\newblock \bibinfo{title}{Large anomalous {Hall} effect in a non-collinear
  antiferromagnet at room temperature}.
\newblock \emph{\bibinfo{journal}{Nature}} \textbf{\bibinfo{volume}{527}},
  \bibinfo{pages}{212--215} (\bibinfo{year}{2015}).

\bibitem{Wang2018}
\bibinfo{author}{Wang, Q.} \emph{et~al.}
\newblock \bibinfo{title}{Large intrinsic anomalous {Hall} effect in
  half-metallic ferromagnet {Co$_3$Sn$_2$S$_2$} with magnetic {Weyl} fermions}.
\newblock \emph{\bibinfo{journal}{Nat. Commun.}} \textbf{\bibinfo{volume}{9}},
  \bibinfo{pages}{1--8} (\bibinfo{year}{2018}).

\bibitem{Liu2019}
\bibinfo{author}{Liu, D.~F.} \emph{et~al.}
\newblock \bibinfo{title}{Magnetic {Weyl} semimetal phase in a {Kagom{\'e}}
  crystal}.
\newblock \emph{\bibinfo{journal}{Science}} \textbf{\bibinfo{volume}{365}},
  \bibinfo{pages}{1282--1285} (\bibinfo{year}{2019}).

\bibitem{Zou2019}
\bibinfo{author}{Zou, J.}, \bibinfo{author}{He, Z.} \& \bibinfo{author}{Xu, G.}
\newblock \bibinfo{title}{The study of magnetic topological semimetals by first
  principles calculations}.
\newblock \emph{\bibinfo{journal}{npj Comput. Mater.}}
  \textbf{\bibinfo{volume}{5}}, \bibinfo{pages}{96} (\bibinfo{year}{2019}).

\bibitem{Lee2021}
\bibinfo{author}{Lee, S.~H.} \emph{et~al.}
\newblock \bibinfo{title}{Evidence for a magnetic-field-induced ideal {type-II}
  {Weyl} state in antiferromagnetic topological insulator
  {Mn(Bi$_{1-x}$Sb$_x$)$_2$Te$_4$}}.
\newblock \emph{\bibinfo{journal}{Phys. Rev. X}} \textbf{\bibinfo{volume}{11}},
  \bibinfo{pages}{031032} (\bibinfo{year}{2021}).

\bibitem{Cheng2024}
\bibinfo{author}{Cheng, E.} \emph{et~al.}
\newblock \bibinfo{title}{Tunable positions of {Weyl} nodes via magnetism and
  pressure in the ferromagnetic {Weyl} semimetal {CeAlSi}}.
\newblock \emph{\bibinfo{journal}{Nat. Commun.}} \textbf{\bibinfo{volume}{15}},
  \bibinfo{pages}{1467} (\bibinfo{year}{2024}).

\bibitem{Du2022}
\bibinfo{author}{Du, F.} \emph{et~al.}
\newblock \bibinfo{title}{Consecutive topological phase transitions and
  colossal magnetoresistance in a magnetic topological semimetal}.
\newblock \emph{\bibinfo{journal}{npj Quantum Mater.}}
  \textbf{\bibinfo{volume}{7}}, \bibinfo{pages}{65} (\bibinfo{year}{2022}).

\bibitem{Kuebler2014}
\bibinfo{author}{Kuebler, J.} \& \bibinfo{author}{Felser, C.}
\newblock \bibinfo{title}{Non-collinear antiferromagnets and the anomalous {Hall}
  effect}.
\newblock \emph{\bibinfo{journal}{EPL}} \textbf{\bibinfo{volume}{108}}
  (\bibinfo{year}{2014}).

\bibitem{Nayak2016}
\bibinfo{author}{Nayak, A.~K.} \emph{et~al.}
\newblock \bibinfo{title}{Large anomalous {Hall} effect driven by a
  nonvanishing {Berry} curvature in the noncolinear antiferromagnet
  {Mn$_{3}$Ge}}.
\newblock \emph{\bibinfo{journal}{Sci. Adv.}} \textbf{\bibinfo{volume}{2}},
  \bibinfo{pages}{e1501870} (\bibinfo{year}{2016}).

\bibitem{Ikhlas2017}
\bibinfo{author}{Ikhlas, M.} \emph{et~al.}
\newblock \bibinfo{title}{Large anomalous {Nernst} effect at room temperature in
  a chiral antiferromagnet}.
\newblock \emph{\bibinfo{journal}{Nat. Phys.}} \textbf{\bibinfo{volume}{13}},
  \bibinfo{pages}{1085--1090} (\bibinfo{year}{2017}).

\bibitem{Yang2017}
\bibinfo{author}{Yang, H.} \emph{et~al.}
\newblock \bibinfo{title}{Topological {Weyl} semimetals in the chiral
  antiferromagnetic materials {Mn$_{3}$Ge} and {Mn$_{3}$Sn}}.
\newblock \emph{\bibinfo{journal}{New J. Phys.}} \textbf{\bibinfo{volume}{19}},
  \bibinfo{pages}{015008} (\bibinfo{year}{2017}).

\bibitem{Kuroda2017}
\bibinfo{author}{Kuroda, K.} \emph{et~al.}
\newblock \bibinfo{title}{Evidence for magnetic {Weyl} fermions in a correlated
  metal}.
\newblock \emph{\bibinfo{journal}{Nat. Mater.}} \textbf{\bibinfo{volume}{16}},
  \bibinfo{pages}{1090--1095} (\bibinfo{year}{2017}).

\bibitem{Kimata2019}
\bibinfo{author}{Kimata, M.} \emph{et~al.}
\newblock \bibinfo{title}{Magnetic and magnetic inverse spin {Hall} effects in
  a non-collinear antiferromagnet}.
\newblock \emph{\bibinfo{journal}{Nature}} \textbf{\bibinfo{volume}{565}},
  \bibinfo{pages}{627--630} (\bibinfo{year}{2019}).

\bibitem{Tsai2020}
\bibinfo{author}{Tsai, H.} \emph{et~al.}
\newblock \bibinfo{title}{Electrical manipulation of a topological
  antiferromagnetic state}.
\newblock \emph{\bibinfo{journal}{Nature}} \textbf{\bibinfo{volume}{580}},
  \bibinfo{pages}{608--613} (\bibinfo{year}{2020}).

\bibitem{KREN1970}
\bibinfo{author}{Kren, E.} \& \bibinfo{author}{Kadar, G.}
\newblock \bibinfo{title}{Neutron diffraction study of {Mn$_{3}$Ga}}.
\newblock \emph{\bibinfo{journal}{Solid State Commun.}}
  \textbf{\bibinfo{volume}{8}}, \bibinfo{pages}{1653--1655}
  (\bibinfo{year}{1970}).

\bibitem{NIIDA1983}
\bibinfo{author}{Niida, H.}, \bibinfo{author}{Hori, T.} \&
  \bibinfo{author}{Nakagawa, Y.}
\newblock \bibinfo{title}{Magnetic properties and crystal distortion of
  hexagonal {Mn}$_3${Ga}}.
\newblock \emph{\bibinfo{journal}{J. Phys. Soc. Jpn.}}
  \textbf{\bibinfo{volume}{52}}, \bibinfo{pages}{1512--1514}
  (\bibinfo{year}{1983}).

\bibitem{Boeije2017}
\bibinfo{author}{Boeije, M. F.~J.}, \bibinfo{author}{van Eijck, L.},
  \bibinfo{author}{van Dijk, N.~H.} \& \bibinfo{author}{Br{\"u}ck, E.}
\newblock \bibinfo{title}{Structural and magnetic properties of hexagonal {(Mn,
  Fe)$_{3-\delta}$Ga}}.
\newblock \emph{\bibinfo{journal}{J. Magn. Magn. Mater.}}
  \textbf{\bibinfo{volume}{433}}, \bibinfo{pages}{297--302}
  (\bibinfo{year}{2017}).

\bibitem{Song2021}
\bibinfo{author}{Song, L.} \emph{et~al.}
\newblock \bibinfo{title}{Observation of structural distortion and topological
  {Hall} effect in noncollinear antiferromagnetic hexagonal {Mn$_{3}$Ga}
  magnets}.
\newblock \emph{\bibinfo{journal}{Appl. Phys. Lett.}}
  \textbf{\bibinfo{volume}{119}} (\bibinfo{year}{2021}).

\bibitem{Liu2017}
\bibinfo{author}{Liu, Z.~H.} \emph{et~al.}
\newblock \bibinfo{title}{Transition from anomalous {Hall} effect to
  topological {Hall} effect in hexagonal non-collinear magnet {Mn$_{3}$Ga}}.
\newblock \emph{\bibinfo{journal}{Sci. Rep.}} \textbf{\bibinfo{volume}{7}},
  \bibinfo{pages}{515} (\bibinfo{year}{2017}).

\bibitem{Zhang2023}
\bibinfo{author}{Zhang, Q.} \emph{et~al.}
\newblock \bibinfo{title}{Polyhedral distortions and unusual magnetic order in
  spinel {FeMn$_{2}$O$_{4}$}}.
\newblock \emph{\bibinfo{journal}{Chem. Mater.}} \textbf{\bibinfo{volume}{35}},
  \bibinfo{pages}{2330--2341} (\bibinfo{year}{2023}).

\bibitem{Yamada1988}
\bibinfo{author}{Yamada, N.}, \bibinfo{author}{Sakai, H.},
  \bibinfo{author}{Mori, H.} \& \bibinfo{author}{Ohoyama, T.}
\newblock \bibinfo{title}{Magnetic properties of
  {$\epsilon$}-{Mn}{$_{3}$}{Ge}}.
\newblock \emph{\bibinfo{journal}{Physica B+C}} \textbf{\bibinfo{volume}{149}},
  \bibinfo{pages}{311--315} (\bibinfo{year}{1988}).

\bibitem{Sukhanov2018}
\bibinfo{author}{Sukhanov, A.~S.} \emph{et~al.}
\newblock \bibinfo{title}{Gradual pressure-induced change in the magnetic
  structure of the noncollinear antiferromagnet {Mn}{$_{3}$}{Ge}}.
\newblock \emph{\bibinfo{journal}{Phys. Rev. B}} \textbf{\bibinfo{volume}{97}},
  \bibinfo{pages}{214402} (\bibinfo{year}{2018}).

\bibitem{Zhang2017}
\bibinfo{author}{Zhang, Y.} \emph{et~al.}
\newblock \bibinfo{title}{Strong anisotropic anomalous {Hall} effect and spin
  {Hall} effect in the chiral antiferromagnetic compounds {Mn$_3$X} ({X= Ge,
  Sn, Ga, Ir, Rh, and Pt})}.
\newblock \emph{\bibinfo{journal}{Phys. Rev. B}} \textbf{\bibinfo{volume}{95}},
  \bibinfo{pages}{075128} (\bibinfo{year}{2017}).

\bibitem{Aroyo2006-1}
\bibinfo{author}{Aroyo, M.~I.} \emph{et~al.}
\newblock \bibinfo{title}{Bilbao crystallographic server: {I}. databases and
  crystallographic computing programs}.
\newblock \emph{\bibinfo{journal}{Z. Kristallogr. Cryst. Mater.}}
  \textbf{\bibinfo{volume}{221}}, \bibinfo{pages}{15--27}
  (\bibinfo{year}{2006}).

\bibitem{Aroyo2006-2}
\bibinfo{author}{Aroyo, M.~I.}, \bibinfo{author}{Kirov, A.},
  \bibinfo{author}{Capillas, C.}, \bibinfo{author}{Perez‑Mato, J.~M.} \&
  \bibinfo{author}{Wondratschek, H.}
\newblock \bibinfo{title}{Bilbao crystallographic server. {II}. representations
  of crystallographic point groups and space groups}.
\newblock \emph{\bibinfo{journal}{Acta Crystallogr. A}}
  \textbf{\bibinfo{volume}{62}}, \bibinfo{pages}{115--128}
  (\bibinfo{year}{2006}).

\bibitem{BilbaoMaxMag}
\bibinfo{author}{Perez-Mato, J.~M.} \emph{et~al.}
\newblock \bibinfo{title}{Symmetry-based computational tools for magnetic
  crystallography}.
\newblock \emph{\bibinfo{journal}{Annu. Rev. Mater. Res.}}
  \textbf{\bibinfo{volume}{45}}, \bibinfo{pages}{217--248}
  (\bibinfo{year}{2015}).

\bibitem{Roth1958}
\bibinfo{author}{Roth, W.~L.}
\newblock \bibinfo{title}{Magnetic structures of {MnO}, {FeO}, {CoO}, and
  {NiO}}.
\newblock \emph{\bibinfo{journal}{Phys. Rev.}} \textbf{\bibinfo{volume}{110}},
  \bibinfo{pages}{1333--1341} (\bibinfo{year}{1958}).

\bibitem{Pomjakushin2024}
\bibinfo{author}{Pomjakushin, V.}
\newblock \bibinfo{title}{On the magnetic and crystal structures of {NiO} and
  {MnO}}.
\newblock \emph{\bibinfo{journal}{Acta Crystallogr. B}}
  \textbf{\bibinfo{volume}{80}} (\bibinfo{year}{2024}).

\bibitem{Hinuma2017}
\bibinfo{author}{Hinuma, Y.}, \bibinfo{author}{Pizzi, G.},
  \bibinfo{author}{Kumagai, Y.}, \bibinfo{author}{Oba, F.} \&
  \bibinfo{author}{Tanaka, I.}
\newblock \bibinfo{title}{Band structure diagram paths based on
  crystallography}.
\newblock \emph{\bibinfo{journal}{Comput. Mater. Sci.}}
  \textbf{\bibinfo{volume}{128}}, \bibinfo{pages}{140--184}
  (\bibinfo{year}{2017}).

\bibitem{Takiguchi2020}
\bibinfo{author}{Takiguchi, K.} \emph{et~al.}
\newblock \bibinfo{title}{Quantum transport evidence of {Weyl} fermions in an
  epitaxial ferromagnetic oxide}.
\newblock \emph{\bibinfo{journal}{Nat. Commun.}} \textbf{\bibinfo{volume}{11}},
  \bibinfo{pages}{4969} (\bibinfo{year}{2020}).

\bibitem{Ali2014}
\bibinfo{author}{Ali, M.~N.} \emph{et~al.}
\newblock \bibinfo{title}{Large, non-saturating magnetoresistance in
  {WTe$_{2}$}}.
\newblock \emph{\bibinfo{journal}{Nature}} \textbf{\bibinfo{volume}{514}},
  \bibinfo{pages}{205--208} (\bibinfo{year}{2014}).

\bibitem{Huang2015}
\bibinfo{author}{Huang, X.} \emph{et~al.}
\newblock \bibinfo{title}{Observation of the chiral-anomaly-induced negative
  magnetoresistance in 3d weyl semimetal {TaAs}}.
\newblock \emph{\bibinfo{journal}{Phys. Rev. X}} \textbf{\bibinfo{volume}{5}},
  \bibinfo{pages}{031023} (\bibinfo{year}{2015}).

\bibitem{Kiyohara2016}
\bibinfo{author}{Kiyohara, N.}, \bibinfo{author}{Tomita, T.} \&
  \bibinfo{author}{Nakatsuji, S.}
\newblock \bibinfo{title}{Giant anomalous {Hall} effect in the chiral
  antiferromagnet {Mn$_{3}$Ge}}.
\newblock \emph{\bibinfo{journal}{Phys. Rev. Appl.}}
  \textbf{\bibinfo{volume}{5}}, \bibinfo{pages}{064009} (\bibinfo{year}{2016}).
 

\bibitem{Gerber2018}
\bibinfo{author}{Gerber, A.}
\newblock \bibinfo{title}{Interpretation of experimental evidence of the
  topological {Hall} effect}.
\newblock \emph{\bibinfo{journal}{Phys. Rev.B}} \textbf{\bibinfo{volume}{98}},
  \bibinfo{pages}{214440} (\bibinfo{year}{2018}).

\bibitem{Kimbell2022}
\bibinfo{author}{Kimbell, G.}, \bibinfo{author}{Kim, C.}, \bibinfo{author}{Wu,
  W.}, \bibinfo{author}{Cuoco, M.} \& \bibinfo{author}{Robinson, J. W.~A.}
\newblock \bibinfo{title}{Challenges in identifying chiral spin textures via
  the topological {Hall} effect}.
\newblock \emph{\bibinfo{journal}{Commun. Mat.}} \textbf{\bibinfo{volume}{3}},
  \bibinfo{pages}{19} (\bibinfo{year}{2022}).

\bibitem{Li2013}
\bibinfo{author}{Li, Y.} \emph{et~al.}
\newblock \bibinfo{title}{Robust formation of skyrmions and topological {Hall}
  effect anomaly in epitaxial thin films of {MnSi}}.
\newblock \emph{\bibinfo{journal}{Phys. Rev. Lett.}}
  \textbf{\bibinfo{volume}{110}}, \bibinfo{pages}{117202}
  (\bibinfo{year}{2013}).

\bibitem{Sticht1989}
\bibinfo{author}{Sticht, J.}, \bibinfo{author}{H{\"o}ck, K.~H.} \&
  \bibinfo{author}{K{\"u}bler, J.}
\newblock \bibinfo{title}{Non-collinear itinerant magnetism: the case of
  {Mn$_{3}$Sn}}.
\newblock \emph{\bibinfo{journal}{J. Phys.: Condens. Matter}}
  \textbf{\bibinfo{volume}{1}}, \bibinfo{pages}{8155} (\bibinfo{year}{1989}).

\bibitem{Brown1990}
\bibinfo{author}{Brown, P.~J.}, \bibinfo{author}{Nunez, V.},
  \bibinfo{author}{Tasset, F.}, \bibinfo{author}{Forsyth, J.~B.} \&
  \bibinfo{author}{Radhakrishna, P.}
\newblock \bibinfo{title}{Determination of the magnetic structure of
  {Mn$_{3}$Sn} using generalized neutron polarization analysis}.
\newblock \emph{\bibinfo{journal}{J. Phys.: Condens. Matter}}
  \textbf{\bibinfo{volume}{2}}, \bibinfo{pages}{9409} (\bibinfo{year}{1990}).

\bibitem{Li2023}
\bibinfo{author}{Li, X.}, \bibinfo{author}{Koo, J.}, \bibinfo{author}{Zhu, Z.},
  \bibinfo{author}{Behnia, K.} \& \bibinfo{author}{Yan, B.}
\newblock \bibinfo{title}{Field-linear anomalous {Hall} effect and {Berry}
  curvature induced by spin chirality in the kagome antiferromagnet
  {Mn$_{3}$Sn}}.
\newblock \emph{\bibinfo{journal}{Nat. Commun.}} \textbf{\bibinfo{volume}{14}},
  \bibinfo{pages}{1642} (\bibinfo{year}{2023}).

\bibitem{Neubauer2009}
\bibinfo{author}{Neubauer, A.} \emph{et~al.}
\newblock \bibinfo{title}{Topological {Hall} effect in the {A} phase of
  {MnSi}}.
\newblock \emph{\bibinfo{journal}{Phys. Rev. Lett.}}
  \textbf{\bibinfo{volume}{102}}, \bibinfo{pages}{186602}
  (\bibinfo{year}{2009}).

\bibitem{Chen2021}
\bibinfo{author}{Chen, T.} \emph{et~al.}
\newblock \bibinfo{title}{Anomalous transport due to weyl fermions in the
  chiral antiferromagnets {Mn$_{3}$X, X= Sn, Ge}}.
\newblock \emph{\bibinfo{journal}{Nat. Commun}} \textbf{\bibinfo{volume}{12}},
  \bibinfo{pages}{572} (\bibinfo{year}{2021}).

\bibitem{Soluyanov2015}
\bibinfo{author}{Soluyanov, A.~A.} \emph{et~al.}
\newblock \bibinfo{title}{{Type-II} {Weyl} {Semimetals}}.
\newblock \emph{\bibinfo{journal}{Nature}} \textbf{\bibinfo{volume}{527}},
  \bibinfo{pages}{495--498} (\bibinfo{year}{2015}).

\bibitem{Ullakko1996}
\bibinfo{author}{Ullakko, K.}, \bibinfo{author}{Huang, J.~K.},
  \bibinfo{author}{Kantner, C.}, \bibinfo{author}{O'Handley, R.~C.} \&
  \bibinfo{author}{Kokorin, V.~V.}
\newblock \bibinfo{title}{Large magnetic-field-induced strains in
  {Ni$_{2}$MnGa} single crystals}.
\newblock \emph{\bibinfo{journal}{Appl. Phys. Lett.}}
  \textbf{\bibinfo{volume}{69}}, \bibinfo{pages}{1966--1968}
  (\bibinfo{year}{1996}).

\bibitem{Krenke2005}
\bibinfo{author}{Krenke, T.} \emph{et~al.}
\newblock \bibinfo{title}{Inverse magnetocaloric effect in ferromagnetic
  {Ni--Mn--Sn} alloys}.
\newblock \emph{\bibinfo{journal}{Nat. Mater.}} \textbf{\bibinfo{volume}{4}},
  \bibinfo{pages}{450--454} (\bibinfo{year}{2005}).

\bibitem{Kouvel1962}
\bibinfo{author}{Kouvel, J.~S.} \& \bibinfo{author}{Hartelius, C.~C.}
\newblock \bibinfo{title}{Anomalous magnetic moments and transformations in the
  ordered alloy {FeRh}}.
\newblock \emph{\bibinfo{journal}{J. Appl. Phys.}}
  \textbf{\bibinfo{volume}{33}}, \bibinfo{pages}{1343--1344}
  (\bibinfo{year}{1962}).

\bibitem{Annaorazov1996}
\bibinfo{author}{Annaorazov, M.~P.}, \bibinfo{author}{Nikitin, S.~A.},
  \bibinfo{author}{Tyurin, A.~L.}, \bibinfo{author}{Asatryan, K.~A.} \&
  \bibinfo{author}{Dovletov, A.~K.}
\newblock \bibinfo{title}{Anomalously high entropy change in {FeRh} alloy}.
\newblock \emph{\bibinfo{journal}{J. Appl. Phys.}}
  \textbf{\bibinfo{volume}{79}}, \bibinfo{pages}{1689--1695}
  (\bibinfo{year}{1996}).

\bibitem{Kato1983}
\bibinfo{author}{Kato, T.}, \bibinfo{author}{Nagai, K.} \&
  \bibinfo{author}{Aisaka, T.}
\newblock \bibinfo{title}{A model of magneto-structural phase transition in
  {MnAs}}.
\newblock \emph{\bibinfo{journal}{J. Phys. C: Solid State Phys.}}
  \textbf{\bibinfo{volume}{16}}, \bibinfo{pages}{3183} (\bibinfo{year}{1983}).

\bibitem{Schiffer1995}
\bibinfo{author}{Schiffer, P.}, \bibinfo{author}{Ramirez, A.~P.},
  \bibinfo{author}{Bao, W.} \& \bibinfo{author}{Cheong, S.-W.}
\newblock \bibinfo{title}{Low temperature magnetoresistance and the magnetic
  phase diagram of {La$_{1-x}$Ca$_{x}$MnO$_{3}$}}.
\newblock \emph{\bibinfo{journal}{Phys. Rev. Lett.}}
  \textbf{\bibinfo{volume}{75}}, \bibinfo{pages}{3336} (\bibinfo{year}{1995}).

\bibitem{Garlea2008}
\bibinfo{author}{Garlea, V.~O.} \emph{et~al.}
\newblock \bibinfo{title}{Magnetic and orbital ordering in the spinel
  {MnV$_{2}$O$_{4}$}}.
\newblock \emph{\bibinfo{journal}{Phys. Rev. Lett.}}
  \textbf{\bibinfo{volume}{100}}, \bibinfo{pages}{066404}
  (\bibinfo{year}{2008}).

\bibitem{Trung2010}
\bibinfo{author}{Trung, N.~T.}, \bibinfo{author}{Zhang, L.},
  \bibinfo{author}{Caron, L.}, \bibinfo{author}{Buschow, K. H.~J.} \&
  \bibinfo{author}{Bruck, E.}
\newblock \bibinfo{title}{Giant magnetocaloric effects by tailoring the phase
  transitions}.
\newblock \emph{\bibinfo{journal}{Appl. Phys. Lett.}}
  \textbf{\bibinfo{volume}{96}}, \bibinfo{pages}{172504}
  (\bibinfo{year}{2010}).

\bibitem{Liu2012}
\bibinfo{author}{Liu, E.} \emph{et~al.}
\newblock \bibinfo{title}{Stable magnetostructural coupling with tunable
  magnetoresponsive effects in hexagonal ferromagnets}.
\newblock \emph{\bibinfo{journal}{Nat. Commun.}} \textbf{\bibinfo{volume}{3}},
  \bibinfo{pages}{873} (\bibinfo{year}{2012}).

\bibitem{Liu2016}
\bibinfo{author}{Liu, J.} \emph{et~al.}
\newblock \bibinfo{title}{Realization of magnetostructural coupling by
  modifying structural transitions in {MnNiSi-CoNiGe} system with a wide
  curie-temperature window}.
\newblock \emph{\bibinfo{journal}{Sci. Rep}} \textbf{\bibinfo{volume}{6}},
  \bibinfo{pages}{23386} (\bibinfo{year}{2016}).

\bibitem{Aryal2017}
\bibinfo{author}{Aryal, A.} \emph{et~al.}
\newblock \bibinfo{title}{Magnetostructural phase transitions and
  magnetocaloric effects in as-cast {Mn$_{1-x}$Al$_{x}$CoGe} compounds}.
\newblock \emph{\bibinfo{journal}{J. Alloys Compd.}}
  \textbf{\bibinfo{volume}{709}}, \bibinfo{pages}{142--146}
  (\bibinfo{year}{2017}).

\bibitem{Czaja2016}
\bibinfo{author}{Czaja, P.} \emph{et~al.}
\newblock \bibinfo{title}{Effect of heat treatment on magnetostructural
  transformations and exchange bias in heusler
  {Ni$_{48}$Mn$_{39.5}$Sn$_{9.5}$Al$_{3}$} ribbons}.
\newblock \emph{\bibinfo{journal}{Acta Mater.}} \textbf{\bibinfo{volume}{103}},
  \bibinfo{pages}{30--45} (\bibinfo{year}{2016}).

\bibitem{Chen2019}
\bibinfo{author}{Chen, J.-H.} \emph{et~al.}
\newblock \bibinfo{title}{Effects of heat treatments on magneto-structural
  phase transitions in {MnNiSi-FeCoGe} alloys}.
\newblock \emph{\bibinfo{journal}{Intermetallics}}
  \textbf{\bibinfo{volume}{112}}, \bibinfo{pages}{106547}
  (\bibinfo{year}{2019}).

\bibitem{Pecharsky2001}
\bibinfo{author}{Pecharsky, V.} \& \bibinfo{author}{Gschneidner, K.~A.}
\newblock \bibinfo{title}{{$Gd_{5}(Si_{x}Ge_{1-x})_{4}$}: An extremum
  material}.
\newblock \emph{\bibinfo{journal}{Adv. Mater.}} \textbf{\bibinfo{volume}{13}},
  \bibinfo{pages}{683--686} (\bibinfo{year}{2001}).

\bibitem{Manosa2010}
\bibinfo{author}{Manosa, L.} \emph{et~al.}
\newblock \bibinfo{title}{Giant solid-state barocaloric effect in the
  {Ni-Mn-In} magnetic shape-memory alloy}.
\newblock \emph{\bibinfo{journal}{Nat. Mater.}} \textbf{\bibinfo{volume}{9}},
  \bibinfo{pages}{478--481} (\bibinfo{year}{2010}).

\bibitem{Daeweritz2006}
\bibinfo{author}{Daeweritz, L.}
\newblock \bibinfo{title}{Interplay of stress and magnetic properties in
  epitaxial {MnAs} films}.
\newblock \emph{\bibinfo{journal}{Rep. Prog. Phys.}}
  \textbf{\bibinfo{volume}{69}}, \bibinfo{pages}{2581--2629}
  (\bibinfo{year}{2006}).

\bibitem{GSAS2}
\bibinfo{author}{Toby, B.~H.} \& \bibinfo{author}{Von~Dreele, R.~B.}
\newblock \bibinfo{title}{{GSAS-II}: the genesis of a modern open-source all
  purpose crystallography software package}.
\newblock \emph{\bibinfo{journal}{J. Appl. Crystallogr.}}
  \textbf{\bibinfo{volume}{46}}, \bibinfo{pages}{544--549}
  (\bibinfo{year}{2013}).

\bibitem{Campbell2006}
\bibinfo{author}{Campbell, B.~J.}, \bibinfo{author}{Stokes, H.~T.},
  \bibinfo{author}{Tanner, D.~E.} \& \bibinfo{author}{Hatch, D.~M.}
\newblock \bibinfo{title}{{ISODISPLACE}: a web-based tool for exploring
  structural distortions}.
\newblock \emph{\bibinfo{journal}{J. Appl. Crystallogr.}}
  \textbf{\bibinfo{volume}{39}}, \bibinfo{pages}{607--614}
  (\bibinfo{year}{2006}).

\bibitem{Blochl1994}
\bibinfo{author}{Bl\"ochl, P.~E.}
\newblock \bibinfo{title}{Projector augmented-wave method}.
\newblock \emph{\bibinfo{journal}{Phys. Rev. B}} \textbf{\bibinfo{volume}{50}},
  \bibinfo{pages}{17953--17979} (\bibinfo{year}{1994}).

\bibitem{Kresse1999}
\bibinfo{author}{Kresse, G.} \& \bibinfo{author}{Joubert, D.}
\newblock \bibinfo{title}{From ultrasoft pseudopotentials to the projector
  augmented-wave method}.
\newblock \emph{\bibinfo{journal}{Phys. Rev. B}} \textbf{\bibinfo{volume}{59}},
  \bibinfo{pages}{1758--1775} (\bibinfo{year}{1999}).

\bibitem{Perdew1996}
\bibinfo{author}{Perdew, J.~P.}, \bibinfo{author}{Burke, K.} \&
  \bibinfo{author}{Ernzerhof, M.}
\newblock \bibinfo{title}{Generalized gradient approximation made simple}.
\newblock \emph{\bibinfo{journal}{Phys. Rev. Lett.}}
  \textbf{\bibinfo{volume}{77}}, \bibinfo{pages}{3865--3868}
  (\bibinfo{year}{1996}).

\bibitem{Kresse1996a}
\bibinfo{author}{Kresse, G.} \& \bibinfo{author}{Furthm\"uller, J.}
\newblock \bibinfo{title}{Efficient iterative schemes for ab initio
  total-energy calculations using a plane-wave basis set}.
\newblock \emph{\bibinfo{journal}{Phys. Rev. B}} \textbf{\bibinfo{volume}{54}},
  \bibinfo{pages}{11169--11186} (\bibinfo{year}{1996}).

\bibitem{Kresse1996b}
\bibinfo{author}{Kresse, G.} \& \bibinfo{author}{Furthm{\"u}ller, J.}
\newblock \bibinfo{title}{Efficiency of ab-initio total energy calculations for
  metals and semiconductors using a plane-wave basis set}.
\newblock \emph{\bibinfo{journal}{Comput. Mater. Sci.}}
  \textbf{\bibinfo{volume}{6}}, \bibinfo{pages}{15--50} (\bibinfo{year}{1996}).

\bibitem{Pizzi2020}
\bibinfo{author}{Pizzi, G.} \emph{et~al.}
\newblock \bibinfo{title}{{Wannier90} as a community code: new features and
  applications}.
\newblock \emph{\bibinfo{journal}{J. Phys.: Condens. Matter}}
  \textbf{\bibinfo{volume}{32}}, \bibinfo{pages}{165902}
  (\bibinfo{year}{2020}).

\bibitem{Zhi2022}
\bibinfo{author}{Zhi, G.-X.}, \bibinfo{author}{Xu, C.}, \bibinfo{author}{Wu,
  S.-Q.}, \bibinfo{author}{Ning, F.} \& \bibinfo{author}{Cao, C.}
\newblock \bibinfo{title}{{WannSymm}: A symmetry analysis code for wannier
  orbitals}.
\newblock \emph{\bibinfo{journal}{Comput. Phys. Commun.}}
  \textbf{\bibinfo{volume}{271}}, \bibinfo{pages}{108196}
  (\bibinfo{year}{2022}).

\bibitem{Wu2018}
\bibinfo{author}{Wu, Q.}, \bibinfo{author}{Zhang, S.}, \bibinfo{author}{Song,
  H.-F.}, \bibinfo{author}{Troyer, M.} \& \bibinfo{author}{Soluyanov, A.~A.}
\newblock \bibinfo{title}{{WannierTools}: An open-source software package for
  novel topological materials}.
\newblock \emph{\bibinfo{journal}{Comput. Phys. Commun.}}
  \textbf{\bibinfo{volume}{224}}, \bibinfo{pages}{405--416}
  (\bibinfo{year}{2018}).

\end{thebibliography}


\section*{Acknowledgments}
This research used resources at the Spallation Neutron Source, a DOE Office of Science User Facility operated by the Oak Ridge National Laboratory. 
The beam time was allocated to POWGEN on proposal number IPTS-24323.1, 24634.1, and IPTS-24773.1. The research by S.O., M.A.M., D.A.T. was sponsored by the Laboratory Directed Research and Development Program (LDRD) of Oak Ridge National Laboratory, managed by UT-Battelle, LLC, for the U.S. Department of Energy (Project No. 9533), with later stage supported by the U.S. DOE, Office of Science, Basic Energy Sciences, Materials Sciences and Engineering Division (S.O.), and U.S. Department of Energy, Office of Science, National Quantum Information Science Research Centers, Quantum Science Center (M.A.M. and D.A.T.).
This research used resources of the Compute and Data Environment for Science (CADES) at the Oak Ridge National Laboratory, which is supported by the Office of Science of the U.S. Department of Energy under Contract No. DE-AC05-00OR22725.

\section*{Author Contributions}
Q.Z. initialized this project. T.-H.Y., D.A.T. and Q.Z. carried out the neutron experiments. S.O. performed the DFT calculations. M.A.M prepared sample and conducted x-ray diffraction, magnetization and transport measurements. T.-H.Y. and Q.Z. analyzed neutron and transport data. T.-H.Y., S.O. and Q.Z. wrote the paper with comments from all the authors.

\section*{Competing interests}
There are no competing interests to declare.

\newpage
\begin{table}[tb]
\centering
\caption{\textbf{Refined atomic positions and magnetic moments at 350~K (represented in a hexagonal cell).} The refinement was performed in the magnetic space group \magsgh{}, with lattice parameters $a_\text{H}=b_\text{H}=$5.4078(9)~\AA, $c_\text{H}=$4.3568(2)~\AA.} 
\vspace{6pt}
\begin{tabular}{lccccccccc}
\hline
\textbf{Atom} & \textbf{x} & \textbf{y} & \textbf{z} & \textbf{Symm. constr.} & $M_x$ & $M_y$ & $M_z$ & $|\mathbf{M}|$ \\
\hline
Mn(1) & 0.6722(3) & 0.8361(5) & 0.2500 & $m_x$, $-m_x$, 0 & –1.18(17) & 1.18(17) & 0.00 & 2.06(25) \\
Mn(2) & 0.8361(7) & 0.6722(3) & 0.7500 & $m_x$, $m_y$, 0 & –1.18(17) & -2.36(17) & 0.00 & 2.06(25) \\
Ga & 0.3333 & 0.6667 & 0.7500 & --- & --- & --- & --- & --- \\
\hline
\end{tabular}
\label{tab:MagStrTN1}
\end{table}

\begin{table}[tb]
\centering
\caption{\textbf{Refined atomic positions and magnetic moments at 30~K.} The refinement was performed in the magnetic space group \magsgl{}, with lattice parameters $a_\text{M}=$5.4281(1)~\AA, $b_\text{M}=$4.3396(1)~\AA, $c_\text{M}=$5.3223(7)~\AA, and $\beta=$119.45(7)$^{\circ{}}$.}
\vspace{6pt}
\begin{tabular}{lccccccccc}
\hline
\textbf{Atom} & \textbf{x} & \textbf{y} & \textbf{z} & \textbf{Symm. constr.} & $M_x$ & $M_y$ & $M_z$ & $|\mathbf{M}|$ \\
\hline
Mn(1) & 0.3433(2) & 0.2500 & 0.1766(1) & $m_x$, 0, $m_z$ &  1.59(16) & 0.00 &  2.93(29) & 2.56(24) \\
Mn(2) & 0.8423(5) & 0.2500 & 0.1771(9) & $m_x$, 0, $m_z$ &  2.54(21) & 0.00 & –0.045(18) & 2.56(28) \\
Mn(3) & 0.8439(5) & 0.2500 & 0.6782(1) & $m_x$, 0, $m_z$ & –2.75(23) & 0.00 & –0.43(16) & 2.56(30) \\
Ga & 0.3388(4) & 0.2500 & 0.6677(6) & --- & --- & --- & --- & --- \\
\hline
\end{tabular}
\label{tab:MagStrTN2}
\end{table}

\newpage

\clearpage
\appendix 
\pagenumbering{arabic} 
\renewcommand{\thefigure}{\arabic{figure}}
\renewcommand{\thetable}{\arabic{table}}
\renewcommand{\theequation}{\arabic{equation}}
\renewcommand{\thepage}{\arabic{page}}
\setcounter{figure}{0}
\setcounter{table}{0}
\setcounter{equation}{0}
\setcounter{page}{1} 

\title{\LARGE \textbf{Supplementary Information}}
\maketitle
\subsection*{DFT of magnetic structures in the low-temperature 
monoclinic phase}
Here, we provide the theoretical detail of the investigation of the ground-state magnetic ordering with the low-temperature monoclinic structure. 
As described in the main text, we carry out density functional theory (DFT) calculations. 
We consider nine initial spin configurations, that are consistent with the lattice symmetry. 
These configurations are grouped as shown in Fig.~\ref{fig:SI_DFT_initial}:
the first three and second three orderings exhibit 120$^{\circ}$-like spin arrangements but with opposite chirality, 
and the last three orderings feature nearly-collinear arrangements. 
We allowed these initial states to relax to stable configurations. 
The resulting magnetic orderings are summarized in Fig.~\ref{fig:SI_DFT} sorted by the ascending order of total energy. 
The same labels (A-I) used for the initial states were retained.
The lowest-energy orderings have the same spin chirality as that with high-temperature hexagonal structure as well as that in Mn$_3$Sn and Mn$_3$Ge, but their spin orientations are modified due to the lower crystal symmetry. 
The magnetic ordering labeled \textbf{A} is consistent with our neutron diffraction measurements.

\begin{figure}
    \centering
    \includegraphics[width=0.6\linewidth]{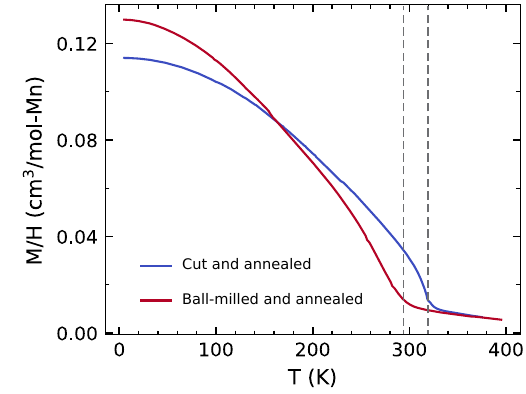}
\caption{
\textbf{Ball milling effect on the locations of $T_\text{N2}$.}
Both originated from the same cast and annealed boule of \MnGa{}. The red curve represents the sample that was ball milled and annealed for 17 hours ($T_\text{N2}=$~295~K). The blue curve represents the sample that was cut from the boule, and annealed directly for 17 hours without prior ball milling process ($T_\text{N2}=$~320~K). The data were measured in an applied field of 10~kOe.
}
\label{fig:SI_BM_XRD}
\end{figure}
\newpage

\begin{figure}
    \centering
    \includegraphics[width=1.0\linewidth]{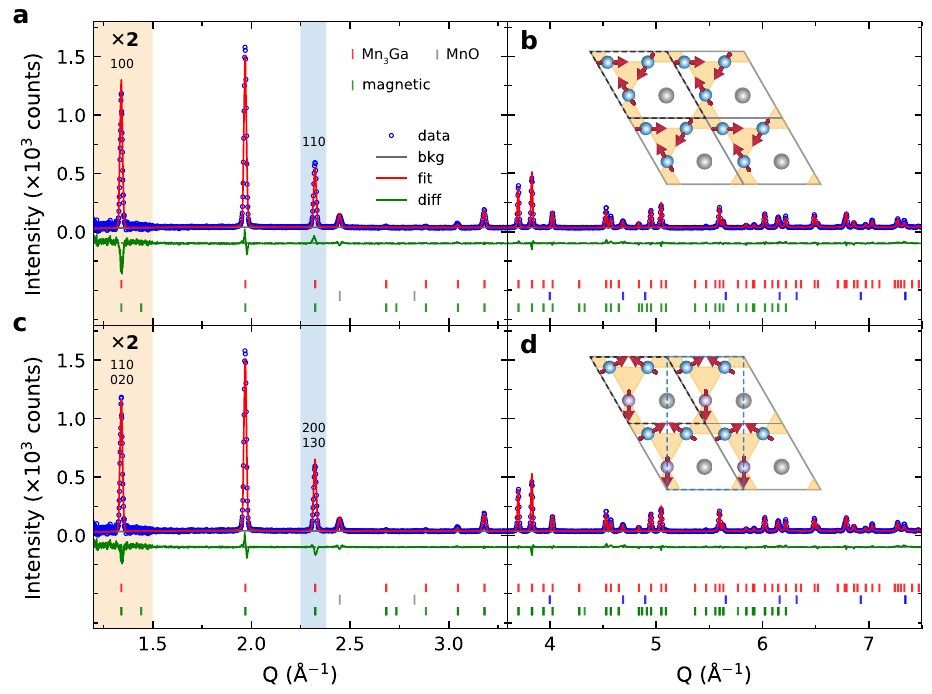}
\caption{
\textbf{Rietveld refinements of neutron powder diffraction patterns at 350~K.} \textbf{a}, \textbf{b} Fits obtained using the hexagonal magnetic model (\magsghrpt{}) and \textbf{c}, \textbf{d} the orthorhombic magnetic model (\magsgh{}). Low–$Q$ regions \textbf{a}, \textbf{c} highlight the magnetic scattering, where the orthorhombic model provides a substantially improved refinement of the data and peak ratio $I(100)_{H}/I(110)_{H}$, as emphasized in the shaded regions. The yellow shaded areas are magnified by a factor of 2 for clarity. High–$Q$ regions \textbf{b}, \textbf{d} show comparable agreement for both models, indicating an equivalent description of the nuclear lattice contribution.
}
\label{fig:SI_350KMagStr}
\end{figure}
\newpage

\begin{figure}
    \centering
    \includegraphics[width=1.0\linewidth]{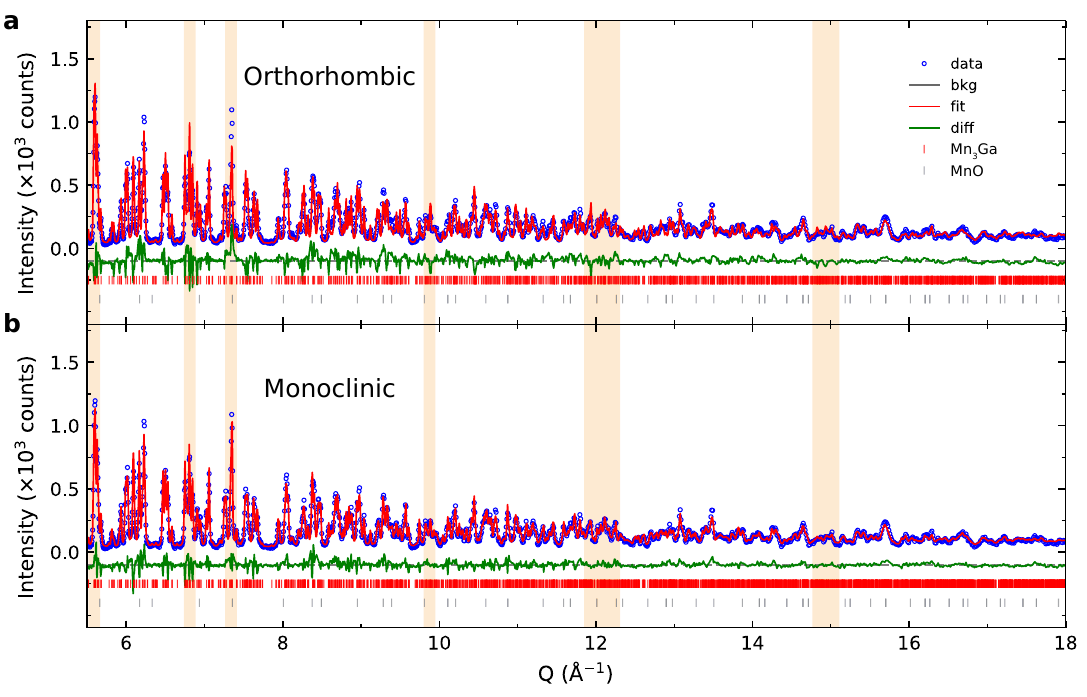}
\caption{
\textbf{Rietveld refinements of high-$Q$ neutron powder diffraction data with negligible magnetic contribution at 200~K.} \textbf{a} Orthorhombic ($Cmcm$) and \textbf{b} monoclinic ($P2_{1}/m$) structural models. 
The monoclinic model provides a better fit ($R_w = 0.0716$) than the orthorhombic model ($R_w = 0.0915$), particularly in reproducing the shapes and intensities of Bragg peaks. 
Shaded regions indicate areas where the monoclinic model shows noticeable improved agreement.
}
\label{fig:SI_NPD}
\end{figure}
\newpage

\begin{figure}
    \centering
    \includegraphics[width=0.6\linewidth]{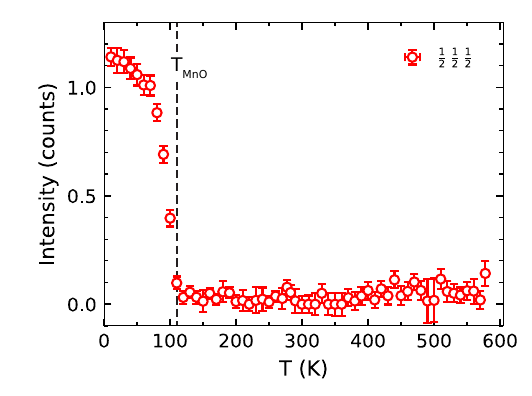}
\caption{
\textbf{Temperature dependence of the $\frac{1}{2}\ \frac{1}{2}\ \frac{1}{2}$ magnetic Bragg peak intensity in \MnO{}.} 
The peak intensity was extracted by fitting with a pseudo-Voigt function using a fixed mixing factor and full width at half maximum for each temperature. 
The magnetic transition temperature near 110~K is consistent with previous reports.
}
\label{fig:SI_MnO}
\end{figure}
\newpage

\begin{figure}
    \centering
    \includegraphics[width=0.8\linewidth]{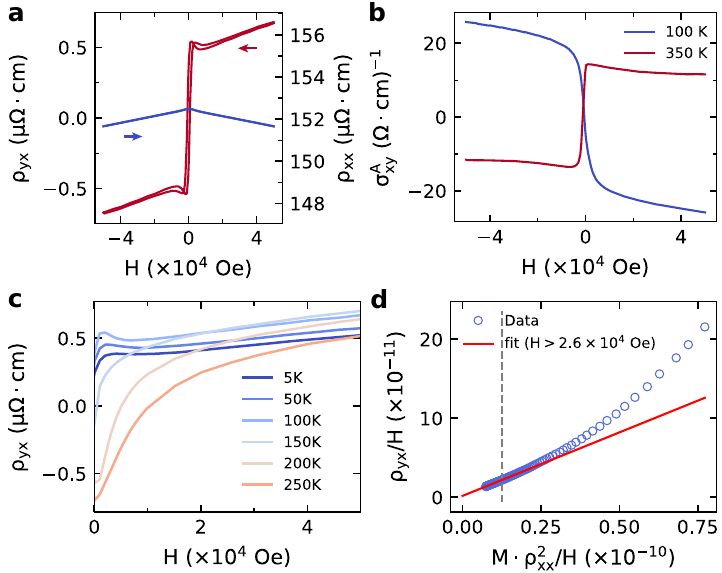}
\caption{
\textbf{Field-dependent Hall resistivity $\rho_{yx}$, longitudinal resistivity $\rho_{xx}$, and anomalous Hall conductivity $\sigma_{xy}^{A}$ component.}
\textbf{a} Field dependence of $\rho_{yx}(H)$ and $\rho_{xx}(H)$ at 100~K. The longitudinal resistivity is significantly larger than the Hall resistivity over the entire field range. 
\textbf{b} Anomalous Hall conductivity at 100~K and 350~K. 
\textbf{c} Expanded view of $\rho_{yx}(H)$ in the vicinity of $T_\text{N2}$, highlighting the evolution across the transition.
\textbf{d} Scaling analysis used to extract the ordinary Hall effect (OHE) and anomalous Hall effect (AHE) contributions. 
}
\label{fig:SI_Hall}
\end{figure}
\newpage


\begin{figure}
    \centering
    \includegraphics[width=0.8\linewidth]{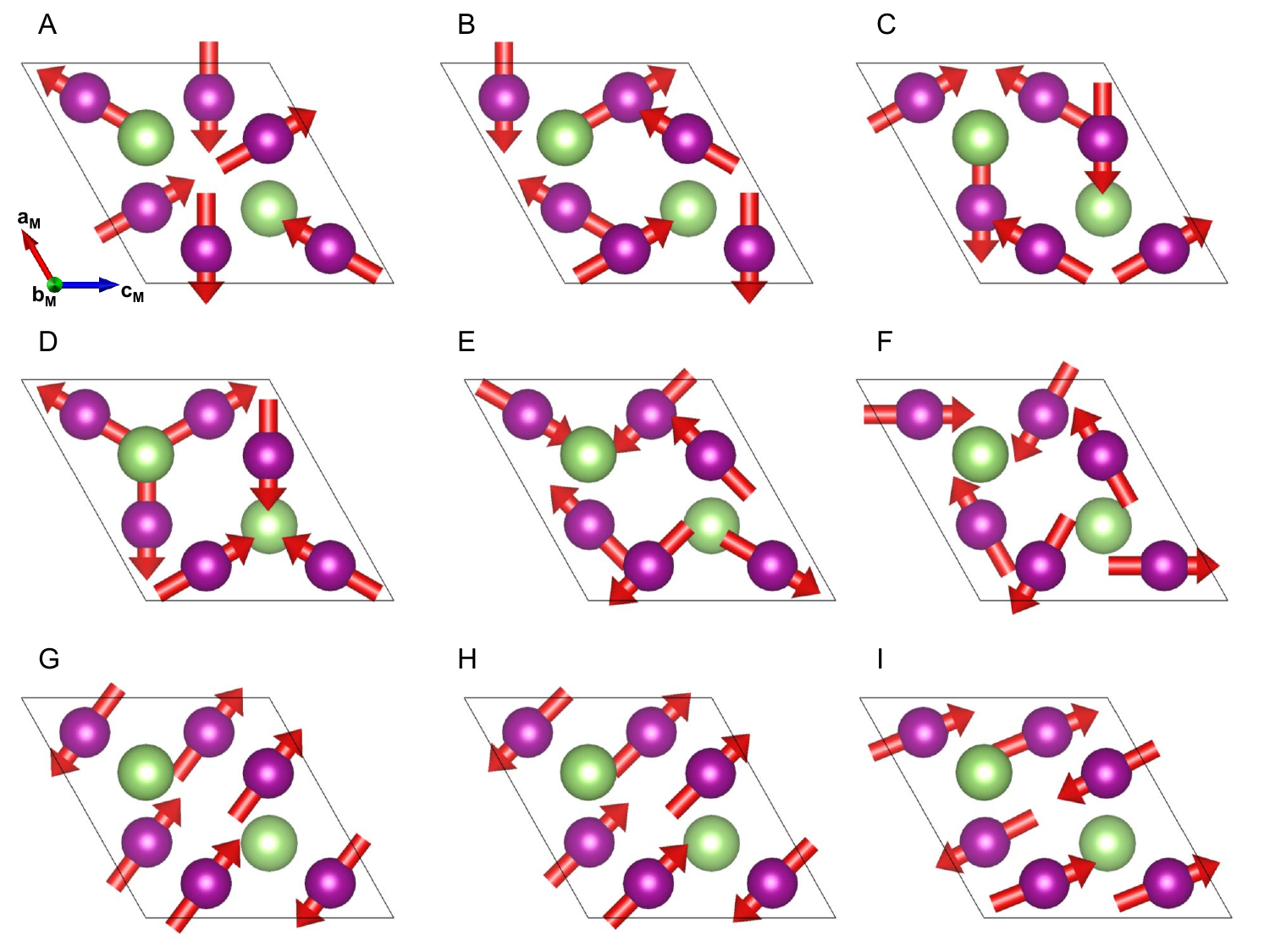}
\caption{
\textbf{Initial magnetic configurations used in DFT calculation 
for the low-temperature monoclinic structure. }
Magnetic configurations (A--C) and (D--F) have 120$^{\circ}$-like spin arrangements with opposite chirality, while 
(G--I) have nearly-collinear arrangements. 
}
\label{fig:SI_DFT_initial}
\end{figure}

\begin{figure}
    \centering
    \includegraphics[width=0.8\linewidth]{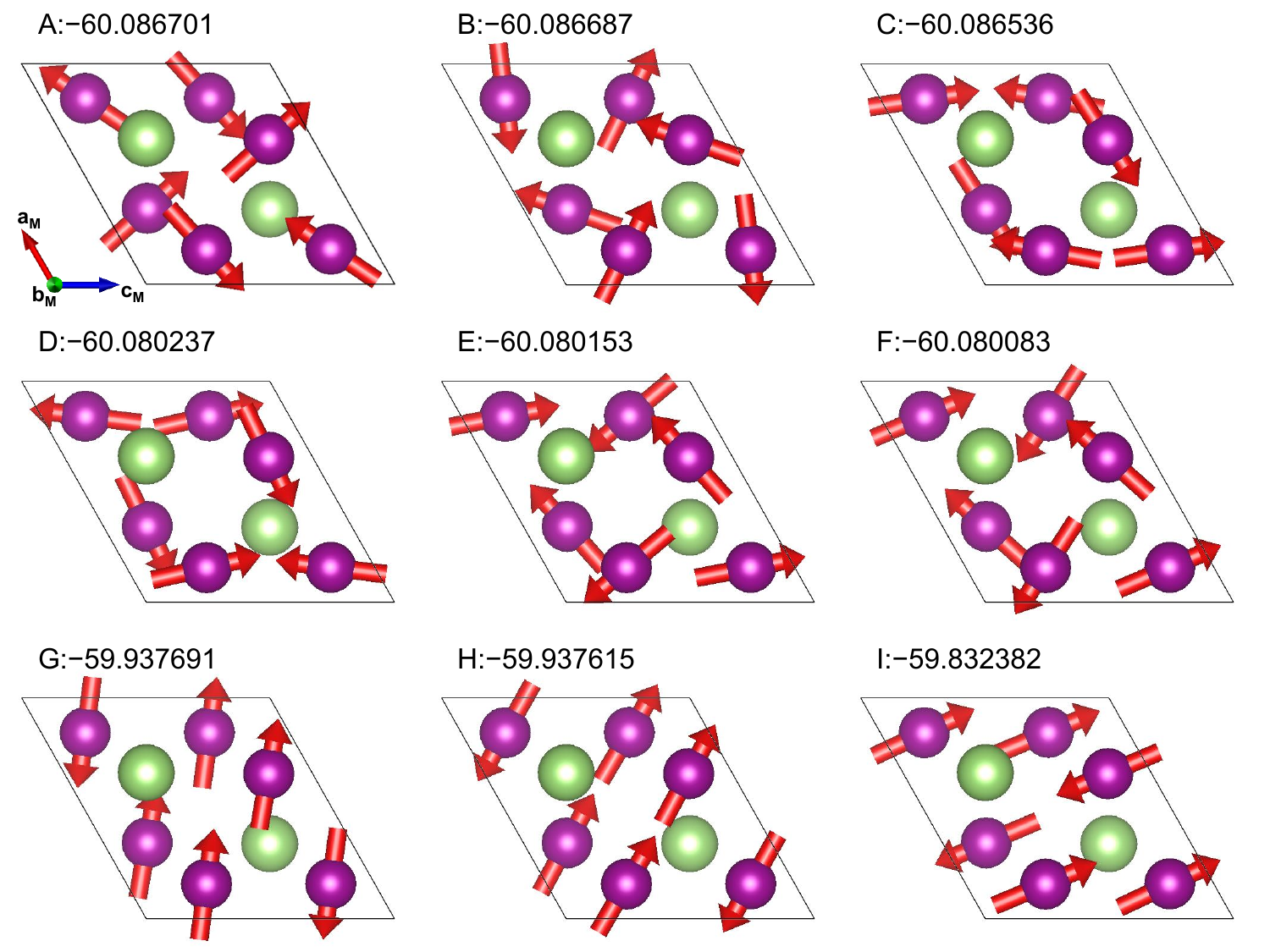}
\caption{
\textbf{DFT calculation results of the total energy (in eV per unit cell), arranged in ascending order from A to I, for nine magnetic configurations of the low-temperature monoclinic structure.} The magnetic configuration designated as A aligns with the results obtained from the Rietveld analysis of neutron diffraction data.
}
\label{fig:SI_DFT}
\end{figure}

\end{document}